# Dynamics of the internal gravity waves in the heterogeneous and nonstationary stratified mediums

**Vitaly V. Bulatov and Yuriy V. Vladimirov**

**Institute for Problems in Mechanics
Russian Academy of Sciences
Pr.Vernadskogo 101-1, 117526 Moscow, Russia**

**bulatov@index-xx.ru**

**Abstract**

*In the present paper in the assumption of the slowness of variation of the vertically stratified medium parameters in the horizontal direction within the time we have analyzed the evolution of the non-harmonic wave trains of the internal gravity waves. The particular form of the wave train can be expressed through some special functions, for example, Airy functions, Fresnel integrals, Pearsy integrals, etc., and is determined by the local behavior of the dispersion curves of the separate modes near to the corresponding singular points. The solution of this problem is possible using the modified version of the space-time ray-tracing method offered by the authors (the method of the ray optics), the fundamental difference of which consists, that the asymptotic notation of such a solution should be searched for in the form of the series using the non-integral degrees of some small parameter, the asymptotic forms of the solution at analysis of evolution of the non-harmonic wave trains present in the stratified non-stationary horizontally-non-uniform mediums is searched in the form of the series using the non-integral degrees of some small parameter, at that the exponent depends on the concrete type of the wave train notation. The particular form of the notation is determined from the asymptotical behavior of the solution in the stationary horizontally-homogeneous event. The phase of the wave train will be determined from the corresponding eikonal equation, which can be solved numerically using the characteristics (rays). The amplitude of the wave train is determined from some law of preservation along the characteristics (rays).*

**Introduction**

As is well known, the horizontal inhomogeneity and nonstationarity exert influence on the stratified natural mediums (the ocean, the atmosphere). To the number of the most specific typical horizontal inhomogeneities of the real ocean it is possible to refer the variations of the shape of the bottom, inhomogeneity of the field of density and changeability of the average flows. The precise value of the analytical solution of the obtained problem, for example by the method of separation of variables, may be gained only in the event, if the density distribution and the shape of the bottom are described by the simple enough model functions. When the shape of the bottom and stratification are arbitrary, then it is possible only to build the asymptotic notation of the solution of the problem in the near and far zones, however for description of the field of the internal waves between these zones it is necessary to have the precise numerical solution of the problem.

If the depth of the ocean and its density varies slowly, as compared with the characteristic length of the internal waves, that is rather well fulfilled at the real ocean, then at searching the solution of the problem of propagation of the internal waves above the variable shape of the bottom it is possible to use the method of geometrical optics. In the present paper the uniform asymptotic forms of the far (distant) field of the internal waves for the case of the constant depth have been developed. It was demonstrated, that the far field represents the sum of the modes, each of which is inside the Mach cone and the asymptotics of each mode is expressed through Airy functions (Airy waves) or Fresnel integrals ( Fresnel waves).

In the present paper in the assumption of the slowness of variation of the vertically stratified medium parameters in the horizontal direction within the time we have analyzed the evolution of the non-harmonic wave trains of the internal gravity waves. The particular form of the wave train can be expressed through some special functions, for example, Airy functions, Fresnel integrals, Pearsy integrals, etc., and is determined by the local behavior of the dispersion curves of the separate modes near to the corresponding singular points.

The solution of this problem is possible using the modified version of the space-time ray-tracing method offered by the authors (the method of the ray optics), the fundamental difference of which consists, that the asymptotic notation of such a solution should be searched for in the form of the series using the non-integral degrees of some small parameter, the asymptotic forms of the solution at analysis of evolution of the non-harmonic wave trains present in the stratified non-stationary horizontally-non-uniform mediums is searched in the form of the series using the non-integral degrees of some small parameter, at that the exponent depends on the concrete type of the wave train notation.

The particular form of the notation is determined from the asymptotical behavior of the solution in the stationary horizontally-homogeneous event. The phase of the wave train will be determined from the corresponding eikonal equation, which can be solved numerically using the characteristics (rays). The amplitude of the wave train is determined from some law of preservation along the characteristics (rays).

At analysis of the wave trains of the internal waves in the stratified mediums with the slow-changing parameters usually it is supposed, that such a wave train is locally harmonic. Unlike the majority of the publications devoted to this problem study, the modified method of the ray optics developed by the authors in the form of some special functions expansion enables to describe the structure of the wave fields near to and far from the wavefronts. The problems concerning the evolution of the non-harmonic wave trains of the internal gravity waves in the layer of the arbitrarily stratified medium of the variable depth with the non-stationary and non-uniform in the horizontal direction density.

Using the asymptotic notations of the wave field at the great spacing intervals from the source for the event of the constant bottom, the problem to construct the uniform asymptotics of the far field for the case of the slow-varying bottom is solved by one of the modified method of the ray optics - by the method of " the vertical modes - the horizontal rays " (within the given method of the slowness, the variations in the vertical direction are not supposed).

The gained solutions represent expansions of the special type waves - Airy or Fresnel waves, and describe not only the evolution of these non-harmonic wave trains at their propagation above the slow-varying bottom, but also the structure of the wave field of each separate mode both near and far from the wavefronts of the separate modes.

The independent variable of Airy or Fresnel waves is determined from the solution of the corresponding eikonal equation. The amplitude of the wave field at that is determined from the law of conservation of energy along the ray tube.

For the model distributions of the shape of the bottom and stratification describing the typical structure of the ocean shelf we have developed the exact analytical expressions for the rays and the analyzed singularities of the phase structure of the wave field. In particular it was shown, that depending on the path of the source motion, the shape of the bottom, and stratifications there may appear the different singularities of the structure of the wave fields. We have analyzed the effect of the spatial " shutting " concerning the low-frequency components of the wave field, which is generated at the source movement at supercritical speed along the ocean shore.

Similarly we have built the asymptotic notations of the solutions on propagation of the non-harmonic wave trains in the heterogeneous in the horizontal direction medium and in the non-stationary density. The conducted.numerical calculations for the typical oceanic parameters indicate the essential influence of the factors of the non-stationarity and the horizontal heterogeneity of the natural parameters on the real dynamics of the internal gravity waves [1-31].

# 1. Basic concepts of the method of the geometric optics

The present section presents the basic provisions of the method of the geometric optics or Wentzel-Kramers-Brillouin (WKB) approximations with consideration of the specificity of the internal gravitational waves. If to consider the internal waves for the case, when the undisturbed field of density $\rho_0(z,x,y)$ depends not only on the depth $z$, but also on the horizontal coordinates $x$ and $y$, then, generally speaking, if the undisturbed density is the function of the horizontal coordinates, then such a density distribution forms some field of the horizontal flows. However these flows are rather slow and as a first approximation they can be neglected, and so it is usually considered, that the field $\rho_0(z,x,y)$ is set a priori and thereby one may suppose the presence of some exterior sources, or the non-conservatism of the system under analysis. It is also quite evident, that, if the internal waves are propagated above the irregular bottom, then such problem does not originate, because the system "the internal wave – the irregular bottom" is conservative and there is no inflow of the energy from the outside.

Let's further consider the linearized system of the equations of hydrodynamics:

$$\rho_0 \frac{\partial \tilde{U}_1}{\partial t} = -\frac{\partial p}{\partial x}$$

$$\rho_0 \frac{\partial \tilde{U}_2}{\partial t} = -\frac{\partial p}{\partial y}$$

$$\rho_0 \frac{\partial \tilde{W}}{\partial t} = -\frac{\partial p}{\partial z} + g\rho \quad (1.1)$$

$$\frac{\partial \tilde{U}_1}{\partial x} + \frac{\partial \tilde{U}_2}{\partial y} + \frac{\partial \tilde{W}}{\partial z} = 0$$

$$\frac{\partial \rho}{\partial t} + \tilde{U}_1 \frac{\partial \rho_0}{\partial x} + \tilde{U}_2 \frac{\partial \rho_0}{\partial y} + \tilde{W} \frac{\partial \rho_0}{\partial z} = 0$$

Here $(\tilde{U}_1, \tilde{U}_2, \tilde{W})$ are the vectors of velocity of the internal gravitational waves, p and $\rho$ are disturbances of the pressure and density, g - an acceleration of gravity ($z$-axis is directed downwards).

Using Boissinesq approximation, which means, that the density $\rho_0(z,x,y)$ in the first three equations of the system (1.1) is considered as a constant value, with the help of the crossover differentiation we shall re-arrange the system (1.1) into:

$$\frac{\partial^4 \tilde{W}}{\partial z^2 \partial t^2} + \Delta \frac{\partial^2 \tilde{W}}{\partial t^2} + \frac{g}{\rho_0} \Delta(\tilde{U}_1 \frac{\partial \rho_0}{\partial x} + \tilde{U}_2 \frac{\partial \rho_0}{\partial y} + \tilde{W} \frac{\partial \rho_0}{\partial z}) = 0$$

$$\frac{\partial}{\partial t}(\Delta \tilde{U}_1 + \frac{\partial^2 \tilde{W}}{\partial z \partial x}) = 0$$

(1.2)

$$\frac{\partial}{\partial t}(\Delta \tilde{U}_2 + \frac{\partial^2 \tilde{W}}{\partial z \partial y}) = 0$$

where $\Delta = \partial^2/\partial x^2 + \partial^2/\partial y^2$.

As the boundary conditions we shall use the condition of "the solid cover"

$$W = 0 \quad \text{at} \quad z = 0, H. \quad (1.3)$$

Let's consider the harmonic waves $(\tilde{U}_1, \tilde{U}_2, \tilde{W}) = e^{i\omega t}(U_1, U_2, W)$ and introduce the dimensionless variables in the formulas

$$x^* = \frac{x}{L}, \quad y^* = \frac{y}{L}, \quad z^* = \frac{z}{h},$$

where L is the characteristic scale of variation of $\rho_0$ in the horizontal direction, h is the characteristic scale of variation of $\rho_0$ in the vertical direction (for example, the width of the thermocline).

In the dimensionless coordinates (1.2) we shall have the following view (the character $*$ further down is omitted):

$$-\omega^2 \left(\frac{\partial^2 W}{\partial z^2} + \varepsilon^2 \Delta W\right) + \varepsilon^2 \frac{g_1}{\rho_0}\left(\varepsilon U_1 \frac{\partial \rho_0}{\partial x} + \varepsilon U_2 \frac{\partial \rho_0}{\partial y} + W \frac{\partial \rho_0}{\partial z}\right) = 0,$$

$$\varepsilon \Delta U_1 + \frac{\partial^2 W}{\partial z \partial x} = 0, \quad \varepsilon \Delta U_2 + \frac{\partial^2 W}{\partial z \partial y} = 0, \tag{1.4}$$

where $\varepsilon = \frac{h}{L} \ll 1, \quad g_1 = \frac{g}{h}.$

The asymptotic solution of (1.4) we shall express in the form, which is typical for the method of the geometrical optics:

$$\mathbf{V}(z,x,y) = \sum_{m=0}^{\infty} (i\varepsilon)^m \mathbf{V}_m(z,x,y) e^{\frac{S(x,y)}{i\varepsilon}}, \tag{1.5}$$

where $\mathbf{V}(z,x,y) = (U_1(z,x,y), U_2(z,x,y), W(z,x,y))$. Functions $S(x,y)$ and $\mathbf{V}_m$, $m = 0,1,...$ are the subject for determination. Later on we shall confine to the search only for the major member of the expansion (1.5) for the vertical component of the velocity $W_0(z,x,y)$. At that from the two last equations of (1.4) we shall have:

$$U_{10} = -\frac{i \partial S/\partial x}{|\nabla S|^2} \frac{\partial W_0}{\partial z}, \tag{1.6}$$

$$U_{20} = -\frac{i \partial S/\partial y}{|\nabla S|^2} \frac{\partial W_0}{\partial z}$$

Where: $|\nabla S| = \left(\frac{\partial S}{\partial x}\right)^2 + \left(\frac{\partial S}{\partial y}\right)^2$.

Insert (1.5) into the first equation of the system (1.4) and equate the terms of O (1) order:

$$\frac{\partial^2 W_0}{\partial z^2} + |\nabla S|^2 \left(\frac{N^2(z,x,y)}{\omega^2} - 1\right) W_0 = 0, \tag{1.7}$$

$$W_0(0,x,y) = W_0(H,x,y) = 0,$$

where $N^2(z,x,y) = \frac{g_1}{\rho_0} \frac{\partial \rho_0}{\partial z}$ - Brunt-Väisälä frequency depending on the horizontal coordinates

For the function $S(x,y)$ we shall have the eikonal equation:

$$\left(\frac{\partial S}{\partial x}\right)^2 + \left(\frac{\partial S}{\partial y}\right)^2 = K^2(x,y) \tag{1.8}$$

The initial conditions for S eikonal equation for a planar case are set on the link: $L: x_0(\alpha), y_0(\alpha)$

$$S(x,y)\big|_L = S_0(\alpha).$$

For the solution of the eikonal equation it is necessary to plot the rays, i.e. the characteristics of the equation (1.8)

$$\frac{dx}{d\sigma} = \frac{p}{K(x,y)}, \qquad \frac{dp}{d\sigma} = \frac{\partial K(x,y)}{\partial x}, \tag{1.9}$$

$$\frac{dy}{d\sigma} = \frac{q}{K(x,y)}, \qquad \frac{dq}{d\sigma} = \frac{\partial K(x,y)}{\partial y}$$

where $p = \partial S/\partial x$, $q = \partial S/\partial y$, $d\sigma$ - a ray length element.
The initial conditions $p_0$ and $q_0$ we shall determine from the system of the equations (1.9)

$$p_0 \frac{\partial x_0}{\partial \alpha} + q_0 \frac{\partial y_0}{\partial \alpha} = \frac{\partial S_0}{\partial \alpha}$$

$$p_0^2 + q_0^2 = K^2(x_0(\alpha), y_0(\alpha))$$

and the initial conditions $x_0(\alpha), y_0(\alpha), p_0(\alpha), q_0(\alpha)$ shall determine the ray $x = x(\sigma, \alpha), y = y(\sigma, \alpha)$. After determination of the beams the eikonal S is being determined by integration over the be

$$S = S_0(\alpha) + \int_0^\sigma K(x(\sigma, \alpha), y(\sigma, \alpha)) d\sigma$$

(1.10)

Now we shall move to determination of the eigenfunction $W_0(z, x, y)$. We would like to underline, that from (1.7) it is possible to determine only the vertical dependence of the eigenfunction $W_0(z, x, y)$. In other words, the eigenfunction $W_0$ may be determined with the accuracy up to the multiplication by the arbitrary function $A_0(x, y)$. Now we shall determine $W_0$ in the form of

$$W_0(z, x, y) = A_0(x, y) \hat{W}_0(z, x, y) \qquad (1.11),$$

where $\hat{W}_0(z, x, y)$ is the solution of the problem (1.7) normalized to the following view

$$\int_0^H (N^2(z, x, y) - \omega^2) \hat{W}_0^2(z, x, y) dz = 1. \qquad (1.12)$$

Let's equate the terms of $O(\varepsilon)$ order after substituting (1.5) into (1.4).

$$\omega^2 \left( \frac{\partial^2 W_1}{\partial z^2} + K^2 \left( \frac{N^2(z, x, y)}{\omega^2} - 1 \right) W_1 \right) = (\omega^2 - N^2)(2\nabla W_0 \nabla S + W_0 \Delta S) + \frac{g_1}{\rho_0} (\nabla S \nabla \rho_0) \frac{\partial W_0}{\partial z} - 2(\nabla N^2 \nabla S) W_0$$

,

$$W_1(0, x, y) = W_1(H, x, y). \qquad (1.13)$$

Further we shall take advantage of the condition of orthogonality of the right part of the equation (1.13) to the function $W_0(z, x, y)$. Multiplying (1.13) by $W_0$ and integrating by z from 0 up to H, we shall receive:

$$\int_0^H (N^2(z, x, y) - \omega^2) \nabla(W_0^2 \nabla S) dz - \frac{g_1}{2} \int_0^H (\nabla S \nabla \ln \rho_0)(\frac{\partial W_0^2}{\partial z}) dz + 2 \int_0^H (\nabla N^2(z, x, y) \nabla S) W_0^2 dz = 0.$$

(1.14)
Transform the second item in (1.14) using integration by parts and the zero boundary conditions for $W_0$:

$$-\frac{g_1}{2} \int_0^H (\nabla S \nabla \ln \rho_0)(\frac{\partial W_0^2}{\partial z}) dz = \frac{\nabla S}{2} \int_0^H \nabla N^2 W_0^2 dz.$$

(1.15)
Transform the first item of (1.14) considering (1.12):

$$\int_0^H (N^2 - \omega^2) \nabla(W_0^2 \nabla S) dz = \nabla(A_0^2 \nabla S) - \int_0^H (\nabla S \nabla N^2) W_0^2 dz.$$

(1.16)

To transform the third item of (1.14), we shall apply the operator of the gradient to the equation (3.1.7), considering $\mathbf{Y} = \nabla W_0$:

$$\frac{\partial^2 \mathbf{Y}(z,x,y)}{\partial z^2} + K^2(x,y)\left(\frac{N^2(z,x,y)}{\omega^2} - 1\right)\mathbf{Y} + W_0 \nabla\left[K^2(x,y)\left(\frac{N^2(z,x,y)}{\omega^2} - 1\right)\right] = 0. \qquad (1.17).$$

Multiplying (1.17) by $W_0$ and integrating by $z$ from 0 up to $H$ and considering (1.12), we shall receive:

$$\int_0^H W_0^2 \nabla N^2(z,x,y) dz = -2A_0^2(x,y) \nabla \ln K(x,y). \qquad (1.18)$$

Finally, we shall rewrite the equation (1.14) using (1.15), (1.16) and (1.18)

$$\nabla A_0^2 \nabla S + A_0^2 \Delta S - 3 \nabla S \nabla \ln K = 0. \qquad (1.19)$$

The transport equation (1.19) we shall solve using characteristics of the eikonal equation (1.9). Using expression for $\Delta S$ along the beams

$$\Delta S = \frac{1}{J}\frac{d}{d\sigma}(JK),$$

where: $J(x,y)$ is the geometrical divergence of the beams. Then we shall transform the transport equation (1.19) to the following law of energy conservation along the rays:

$$\frac{d}{d\sigma}\left(\ln \frac{A_0^2(x,y) J(x,y)}{K^2(x,y)}\right) = 0. \qquad (1.20)$$

The law of the energy conservation (1.20) can be recorded also in the form suitable for finding $A_0$ function:

$$\frac{A_0^2(x(\sigma,\alpha),y(\sigma,\alpha))}{K^2(x(\sigma,\alpha),y(\sigma,\alpha))} da(x(\sigma,\alpha),y(\sigma,\alpha)) = \frac{A_0^2(x_0(\alpha),y_0(\alpha))}{K^2(x_0(\alpha),y_0(\alpha))} da(x_0(\alpha),y_0(\alpha))$$

(1.21),

where: $da(x(\sigma,\alpha),y(\sigma,\alpha)) = J(x(\sigma,\alpha),y(\sigma,\alpha))d\alpha$ is the width of an elementary ray tube. We shall mark, that the current of the wave energy is proportional to $A_0^2 K^{-1} da$, so the (1.21) shows, that in this case the value equal to the current of the wave energy divided by the modulus of the wave vector is kept safe.

For transition to the study of the problem of the non-harmonic wave packets in the smoothly heterogenous in horizontal direction and the non-stationary stratified medium before selection of ansatzs (Anzatz (german.) - a kind of solution), describing propagation of Airy and Fresnel subsurface waves, let's first consider some leading considerations.

Airy surface wave. Inject the slow variables $x^* = \varepsilon x$, $y^* = \varepsilon y$, $t^* = \varepsilon t$ (for $z$ - slowness the variations are not supposed, the character further is neglected), where $\varepsilon = \lambda/L \ll 1$ is the small parameter characterizing the smoothness of the medium variation in the horizontal direction ($\lambda$ - the characteristic length of the wave, $L$ - the scale of the horizontal heterogeneity). Then the system (1.2) with the slow variables will look like this:

$$\frac{\partial^4 W}{\partial z^2 \partial t^2} + \varepsilon^2 \frac{\partial^2 W}{\partial t^2} + \frac{g}{\rho_0}\Delta(\varepsilon U_1 \frac{\partial \rho_0}{\partial x} + \varepsilon U_2 \frac{\partial \rho_0}{\partial y} + W \frac{\partial \rho_0}{\partial z}) = 0 \qquad (1.22)$$

$$\varepsilon \Delta U_1 + \frac{\partial^2 W}{\partial z \partial x} = 0 \quad \varepsilon \Delta U_2 + \frac{\partial^2 W}{\partial z \partial y} = 0 \ .$$

Further, we shall consider the superposition of the cosine waves (with the slow variables $x, y, t$):

$$W = \int \omega \sum_{m=0}^{\infty} (i\varepsilon)^m W_m(\omega, z, x, y) e^{\frac{i}{\varepsilon}[\omega t - S_m(\omega, x, y)]} d\omega. \qquad (3.1.23)$$

Concerning functions $S_m(\omega, x, y)$ it is supposed, that they are odd for $\omega$ and their $\min_\omega \partial S / \partial \omega$ is reached at $\omega = 0$ (for all $x$ and $y$).

Substituting (1.23) in (1.22) it will easy to demonstrate, that the function $W_m(\omega, z, x, y)$ at $\omega = 0$ has a pole of m-th order. Therefore the model integrals $R_m(\sigma)$ for the separate units in (1.23) will be the following expressions:

$$R_m(\sigma) = \frac{1}{2\pi} \int_{-\infty}^{\infty} \left(\frac{i}{\omega}\right)^{m-1} e^{i\left(\frac{\omega^3}{3} - \sigma \omega\right)} d\omega,$$

where the contour of the integration bends around the dot $\omega = 0$ from above, that ensures the exponential fading of the functions $R_m(\sigma)$ at $\sigma \ll 1$. Functions $R_m(\sigma)$ possess the following property:

$$\frac{d R_m(\sigma)}{d\sigma} = R_{m-1}(\sigma),$$

At that $R_0(\sigma) = Ai'(\sigma)$, $R_1(\sigma) = Ai(\sigma)$, $R_2(\sigma) = \int_{-\infty}^{\sigma} Ai(u) du$ etc.

It is obvious, that proceeding from the corresponding properties of Airy integrals, the $R_m(\sigma)$ functions are connected to each other by the following ratios:
$$R_{-1}(\sigma) + \sigma R_1(\sigma) = 0$$
$$R_{-3}(\sigma) + 2 R_0(\sigma) - \sigma^2 R_1(\sigma) = 0.$$

Fresnel wave. As the model integrals $R_m(\sigma)$ describing propagation of Fresnel waves, proceeding from the structure of the solution for the elevation in the horizontally homogeneous case following expressions are used:

$$R_0(\sigma) = \mathrm{Re} \int_0^{\infty} \exp\left(-it\sigma - i\frac{t^2}{2}\right) dt \equiv \mathrm{Re}\, \Phi^*(\sigma) \equiv \Phi(\sigma)$$

It quite evident, that function $\Phi^*(\sigma)$ possesses the following property:

$$\frac{d\Phi^*(\sigma)}{d\sigma} = -\int_0^{\infty} it \exp(-it\sigma - \frac{1}{2}it^2) dt =$$

$$= \int_0^{\infty} \exp(-it\sigma - \frac{1}{2}it^2) d(-it\sigma - \frac{1}{2}it^2) + i\sigma \int_0^{\infty} \exp(-it\sigma - \frac{1}{2}it^2) dt = 1 + i\sigma\, \Phi^*(\sigma)$$

whence, for example, it is possible to obtain:

$$\frac{d^3 \Phi^*(\sigma)}{d\sigma^3} = i\sigma \frac{d^2 \Phi^*(\sigma)}{d\sigma^2} + 2i \frac{d\Phi^*(\sigma)}{d\sigma}$$

or (in terms of functions $R_m(\sigma)$):
$$R_{-1}(\sigma) + i\sigma R_0(\sigma) = 0$$
$$R_{-3}(\sigma) - 2i R_{-1}(\sigma) - i\sigma R_{-2}(\sigma) = 0.$$

Proceeding from the above-stated, and also from the structure of the first item of the uniform asymptotics of Airy and the Fresnel waves in the horizontally uniform medium, the solution of the system (1.22), for example, it is possible to find in the form of (for the separately taken mode $W_n\, \mathbf{U}_n$ with the following exclusion of "n" index):

$$W = \varepsilon^0 W_0(z, x, y, t) R_0(\sigma) + \varepsilon^a W_1(z, x, y, t) R_1(\sigma) + \varepsilon^{2a} W_2(z, x, y, t) R_2(\sigma) + K \qquad (1.24)$$

$$\mathbf{U} = \varepsilon^{1-a}\mathbf{U}_0(z,x,y,t)R_1(\sigma) + \varepsilon \mathbf{U}_1(z,x,y,t)R_2(\sigma) + \varepsilon^{1+a}\mathbf{U}_2(z,x,y,t)R_3(\sigma) + K ,$$

where: argument $\sigma = \left(\dfrac{1}{a}S(x,y,t)\right)^a \varepsilon^{-a}$ .is considered of the order of unity. Expansion (1.24) will be in accord with the general approach of the ray optics method and the space-time ray-tracing method.

Let us also to note, that from the similar structure of the solution follows, that in the heterogeneous in horizontal direction and the non-stationary medium the solution depends both on the "fast" (the vertical coordinate), and the "slow" (the time and horizontal coordinates) variables. Further we shall look for the solution, as a rule, in the "slow" variables. At that those structural members of the solution, which depends on the "fast" variables, are received in the form of the integrals from some slowly-variable functions along the time-space beams. The given method of the solution allows to present the uniform asymptotes of the fields of the internal gravitational waves propagating in the stratified mediums with the slowly-variable parameters, which is true both in the close vicinity and at the far distance from the wave fronts of a separate wave mode.
If it is necessary to describe the behavior of the field only near to the wave front, it is possible to use one of the methods of the ray optics – the method of "traveling wave", and also the low-dispersive approximation in the form of the corresponding local asymptotes and to look for presentation for the argument of the phase functions $\sigma$ in the form of the separate wave mode:

$$\sigma = \alpha(t,x,y)(S(t,x,y) - \varepsilon t)\varepsilon^{-a} ,$$

where the function $S(t,x,y)$ presents the wave front position and is determined from the solution of the eikonal equation

$$\nabla^2 S = c^{-2}(x,y,t)$$

where $c(t,x,y)$ - the maximum group velocity of the corresponding mode, that is the first term of expansion of the dispersing curve in zero. The function $\alpha(t,x,y)$ (the second term of the expansion of the dispersing curve) describes space-time evolution of the pulse width of Airy and Fresnel waves and then will be determined from some energy conservation laws along the characteristics of an eikonal equation, which concrete composition is determined by the physical conditions of the problem.

## 2. Uniform asymptotics of the far field of the internal gravity waves in the layer of the stratified medium with the smoothly varying bottom

In this section we present the built uniform asymptotics of the far field of the internal gravity waves during motion of the point source of mass of the single intensity in the layer of the random stratified liquid with the slowly varying ocean bottom. The gained solutions describe the far field both near to the wave fronts of each separate mode, and far from wave fronts and represent expansions into Airy waves and Fresnel waves, which argument is determined from the solution of the applicable equation of the eikonal. The amplitude of the wave field is determined from the energy conservation law along the ray tube. For the model distributions of the form of the bottom and the stratification, there have been gained the exact analytical notations for rays are gained and the wave field phase structure features have been analyzed.

***Formulation of the problem and selection of the kind of the solution.***

Let the point source of mass moves in the layer $-H(\varepsilon x, \varepsilon y) < z < 0$ ($\varepsilon$ - the small argument) of the stratified liquid with Brunt-Väisälä frequency $N^2(z)$. It is assumed, that the velocity $V$ of the source movement along $x$ axis exceeds the maximal group velocity of propagation of the internal waves, that is the source is moving with the supercritical velocity.

From linearized system of equations of hydrodynamics in Boissinesq approximation we have (see Chapter 1-2)

$$\frac{\partial^2}{\partial t^2}\left(\Delta + \frac{\partial^2}{\partial z^2}\right)w + N^2(z)\,\Delta w = \delta''_{tt}(x+Vt)\,\delta(y)\,\delta'(z-z_0)$$

$$\Delta \mathbf{u} + \nabla \frac{\partial w}{\partial z} = \delta(z-z_0)\,\nabla\left(\delta(x+Vt)\,\delta(y)\right) \quad (2.1)$$

$$\Delta = \frac{\partial^2}{\partial x^2} + \frac{\partial^2}{\partial y^2}, \qquad \nabla = \left(\frac{\partial}{\partial x}, \frac{\partial}{\partial y}\right)$$

Where $\mathbf{u} = (u_1, u_2)$ - the vector of the horizontal velocities, $w$ - the vertical component of velocity, $z_0$ - depth of the source motion.

As the boundary conditions on the surface the "solid cover" condition is used, at the bottom $z = -H(\varepsilon x, \varepsilon y)$ we have the non-flow condition

$$w = 0, \ z = 0$$
$$w = \mathbf{u}\nabla H(\varepsilon x, \varepsilon y), \ z = -H(\varepsilon x, \varepsilon y) \quad (2.2)$$

Solution of the problem (2.1) - (2.2) we shall look for proceeding from the structure of the uniform asymptotics for $H(x, y) = \text{const}$, in the form of the sum of modes, each of which is propagating independently from each other ("adiabatic approximation"} and can be presented in the form of the following asimptotic series:

$$w = A(\varepsilon x, \varepsilon y, z, \varepsilon t)R_0(\sigma) + \varepsilon^a B(\varepsilon x, \varepsilon y, z, \varepsilon t)R_1(\sigma) + K \quad (2.3)$$

$$\mathbf{u} = \mathbf{u}_0(\varepsilon x, \varepsilon y, z, \varepsilon t)R_1(\sigma)\varepsilon^{a-1} + K \quad (2.4)$$

$$R'_{i+1}(\sigma) = R_i(\sigma), \qquad \sigma \equiv \left(\frac{S(\varepsilon x, \varepsilon y, \varepsilon t)}{a\varepsilon}\right)^a$$

Functions $S(\varepsilon x, \varepsilon y, \varepsilon t)$, $A(\varepsilon x, \varepsilon y, z, \varepsilon t)$, $\mathbf{u}_0(\varepsilon x, \varepsilon y, z, \varepsilon t)$ in (2.3), (2.4) are the subject to determiantion. The argument $\sigma(x, y, t)$ is of the order of a unit. The function $R_0(\sigma)$ depending on presence of the homogeneous (non-stratified) sublayer is expressed through Airy function (Airy wave) or through Fresnel integrals (Fresnel wave). Later on without any limitation of generality we shall consider, for example, propagations of Fresnel wave elevation $\eta$ ($w = \partial\eta/\partial t$)

$$R_0(\sigma) = \mathrm{Re}\int_0^\infty \exp\left(-it\sigma - i\frac{t^2}{2}\right)dt \equiv \mathrm{Re}\,\Phi^*(\sigma) \equiv \Phi(\sigma), \quad a = \frac{1}{2}$$

meeting the following ratios:

$$\frac{d\Phi^*(\sigma)}{d\sigma} = 1 + i\sigma\Phi^*(\sigma)$$

$$\frac{d^3\Phi^*(\sigma)}{d\sigma^3} = i\sigma\frac{d^2\Phi^*(\sigma)}{d\sigma^2} + 2i\frac{d\Phi^*(\sigma)}{d\sigma}$$

***Eikonal equation and characteristics. Geometry of the wave field phase structure.***

Now we shall transfer to determination of $S$ function. Substituting (2.3), (2.4) in (2.1) with accuracy up to the terms of the order $\varepsilon^{\frac{3}{2}}$ it is possible to gain

$$\mathbf{u}_0 = -A'_z\sqrt{2S}\,\frac{\nabla S}{\varepsilon\left(\left(\frac{\partial S}{\partial x}\right)^2 + \left(\frac{\partial S}{\partial y}\right)^2\right)} + O(\varepsilon^{\frac{3}{2}})$$

(2.5)

Having substituted (2.3), (2.5) in (2.1), (2.2) and having equated the terms at the equal degrees $\varepsilon$, we shall gain at $\varepsilon^{\frac{1}{2}}$

$$\frac{\partial^2 A}{\partial z^2} + |\mathbf{k}|^2\left(\frac{N^2(z)}{\omega^2} - 1\right)A = 0$$

(2.6)

$$A = 0, \quad z = 0, -H(x, y)$$
$$\mathbf{k} \equiv (p, q) = -\nabla S, \quad \omega = \partial S/\partial t$$

The dispersion dependence designated further through $K(\omega, x, y)$, is determined from the solution of the vertical spectral problem (2.6), $\omega$ - the spectral parameter. Then for determination of function $S$ we have the eikonal equation

$$\left(\frac{\partial S}{\partial x}\right)^2 + \left(\frac{\partial S}{\partial y}\right)^2 = K^2(\omega, x, y)$$

(2.7)

The equation (2.7) is Hamilton-Jacobi equation with the Hamiltonian $|\mathbf{k}|^2 - K^2(\omega, x, y)$ and the characteristic system of this equation looks like

$$\frac{dx}{d\tau} = \frac{p}{K(\omega, x, y) K'_\omega(\omega, x, y)}, \quad \frac{dy}{d\tau} = \frac{q}{K(\omega, x, y) K'_\omega(\omega, x, y)}$$

$$\frac{dp}{d\tau} = \frac{K'_x(\omega, x, y)}{K'_\omega(\omega, x, y)}, \quad \frac{dq}{d\tau} = \frac{K'_y(\omega, x, y)}{K'_\omega(\omega, x, y)}, \quad \frac{d\omega}{d\tau} = 0$$

(2.8)

Having solved the characteristic system (2.8) with the corresponding initial conditions, we gain the two-parameter set of the time-space rays depending on the ray coordinates. At that the choice of the initial conditions and the ray coordinates is determined by the particular kind of the being solved problem. Then the eikonal expressed in ray coordinates $S^*(l, m) \equiv S(x, y, t)$ can be determined by integration along these rays.

As the model instance enabling the analytical solution of the system (2.8) and at the same time qualitatively truly describing the real bottom configuration, we shall consider the following example. Let Brunt-Väisälä frequency is constant $N(z) = N = const$ and the depth of the bottom depends only on one coordinate y linearly $H(y) = \beta y$. Now we shall introduce the coordinate system with x-axis going along the shore $y = 0$.

Let the point source of mass moves from the right to the left in the negative direction of the x axis at the velocity V at the distance $y_0$ offshore and at that in each moment of time t generates the waves of all frequencies within the range $0 < \omega < N$. We shall consider the first mode. Further at $N(z) = const$ from (2.6) we shall have

$$p^2 + q^2 = K^2(\omega, y), \quad K(\omega, y) = \frac{\pi \omega}{H(y)\sqrt{N^2 - \omega^2}}$$

As function $K(\omega, y)$ does not depend on x, then from (2.8) we shall gain

$$\frac{dp}{d\tau} = 0, \quad p = \frac{\omega}{V} = const, \quad q = \pm\sqrt{K^2(\omega, y) - \frac{\omega^2}{V^2}}$$

Then the characteristic system and the initial conditions for the eikonal equation will look like

$$\frac{dx}{d\tau} = \frac{\alpha^2 \gamma^2}{V^2} y^2, \quad \frac{dy}{d\tau} = \pm \alpha^{\frac{3}{2}} \gamma y \sqrt{1 - \frac{\alpha \gamma^2 y^2}{V^2}}$$

$$x = V t_0, \quad t = t_0, \quad y = y_0, \quad t = t_0$$

$$\alpha = 1 - \frac{\omega^2}{N^2}, \quad \gamma = \frac{N\beta}{\pi}$$

(2.9)

Hereinafter the upper sign matches to the field $y > y_0$, the lower sign - to the field $y < y_0$.
Integrating system (2.9), we gain the equations of rays

$$y = \frac{V}{\gamma\sqrt{\alpha}} \frac{1}{\left[\operatorname{ch}\left[\pm\operatorname{arch}\left(\frac{V}{\gamma\sqrt{\alpha}y_0}\right) - \gamma\alpha^{\frac{3}{2}}(t-t_0)\right]\right]}$$

$$x = x_0 \pm y_0\sqrt{\frac{V^2}{\alpha\gamma^2 y_0^2} - 1} - \frac{V}{\gamma\sqrt{\alpha}}\operatorname{th}\left[\pm\operatorname{arch}\left(\frac{V}{\gamma\sqrt{\alpha}y_0}\right) - \gamma\alpha^{\frac{3}{2}}(t-t_0)\right]$$

$$x = x_0 + \frac{\gamma\sqrt{\alpha}}{V} yy_0 \operatorname{sh}(\gamma\alpha^{\frac{3}{2}}(t-t_0))$$

(2.10)

Rays (2.10) as follows from (2.8) are the lines in the space x, y, on which the frequency $\omega$ is constant.

As the maximum of the group velocity of propagation of the internal waves is reached at $\omega = 0$ ($\alpha = 1$), then the wave front position in the moving together with the source coordinate system ($\xi = x + Vt$) is determined from the equation

$$\frac{\partial \xi}{\partial y} = \pm \frac{\sqrt{V^2 - (\gamma y)^2}}{\gamma y}, \quad \xi(y_0) = 0$$

(2.11)

The Solution of (2.11) looks like

$$\xi = \pm \frac{V}{\gamma}(d_1(y) - d_2(y))$$

$$d_1(y) = \operatorname{arch}\left(\frac{V}{\gamma y_0}\right) - \operatorname{arch}\left(\frac{V}{\gamma y}\right)$$

$$d_2(y) = \sqrt{1 - \left(\frac{\gamma y_0}{V}\right)^2} - \sqrt{1 - \left(\frac{\gamma y}{V}\right)^2}$$

Eikonal $S^*$ looks like

$$S^* = \omega(t-t_0) + \int_{t_0}^{t} \frac{K(\omega, y(\tau, t_0, \omega))}{K'_\omega(\omega, y(\tau, t_0, \omega))} d\tau = \omega(t-t_0) - \omega\alpha(t-t_0) = \frac{\omega^3}{N^2}(t-t_0)$$

Then in the coordinate system moving together with the source the lines of the equal phase, i.e. the lines of constant value of then eikonal $S^*$, look like

$$y = \frac{V}{\gamma\sqrt{\alpha}} \frac{1}{\left[ ch\left( \pm arch\left( \frac{V}{\gamma\sqrt{\alpha}y_0} \right) - \gamma\alpha^{\frac{3}{2}}T \right) \right]}$$

$$\xi = VT - \frac{\gamma\sqrt{\alpha}}{V} yy_0 \, sh(\gamma\alpha^{\frac{3}{2}}T), \quad T = \frac{S^*N^2}{\omega^3} \tag{2.12}$$

The Fig..1 presents the results of the calculations for the following parameters, which are close to the parameters of the real shelf: $\beta = 0{,}1$, $N = 0{,}01 c^{-1}$, $y_0 = 1000$ m, $V = 2N\beta y_0 / \pi$.

Presented on the Fig..1 the envelope of the set of the rays, caustic surface, is the geometrical place of the points $\xi = \xi(y)$, satisfying (2.12) and the condition $\xi'(y) = 0$. Therefore, Fresnel wave with the concrete value of the phase $S^*$ has the corresponding critical values with respect to the variable $y$, limiting its field of propagation from the source motion traverse. As one can see from the presented numerical results, with the increase of the phase $S^*$ the rays penetration zone is expanding (with respect to the variable $y$).

Thus, with the increasing distance $y$ from the traverse of the source motion in the coming wave field the share of the low-frequency components, that is Fresnel waves with the small values of the phase is diminishing (the group velocity is diminishing with the frequency growth) and with the chosen dependence of the bottom configuration (the group velocity grows with removal from the shore $y=0$) in this connection rather high-frequency field components with the lower group velocity have possibility to propagate to the greater distances with respect to $y$ from the shore, than the low frequency components.

***Energy conservation law and determination of the amplitude dependence.***

For determination of the amplitude dependence $A$ we shall substitute (2.3), (2.4) into (2.1), (2.2) and equate terms of the order $\varepsilon^{\frac{3}{2}}$. As a result, having taken advantage of the properties of Fresnel integrals, we shall gain

$$\omega^2 B''_{zz} + |k|^2 (N^2(z) - \omega^2)B = (N^2(z) - \omega^2)(2\nabla S \nabla^* A + A\Delta S) +$$

$$+ 2\omega \left( \frac{\partial^*}{\partial t} A''_{zz} - |k|^2 \frac{\partial^* A}{\partial t} \right) + \frac{\partial^* \omega}{\partial t}(A''_{zz} - |k|^2 A) - 4\omega A \nabla S \nabla^* \omega \tag{2.13}$$

$$B = 0, \quad z = 0 \tag{2.14}$$

$$B = 2\nabla S \nabla H A'_z \sqrt{S\left( \left( \frac{\partial S}{\partial x} \right)^2 + \left( \frac{\partial S}{\partial y} \right)^2 \right)^{-1}}, \quad z = -H(x,y)$$

$$\nabla^* = \nabla + \frac{\partial}{\partial \omega} \nabla \omega, \quad \frac{\partial^*}{\partial t} = \frac{\partial}{\partial t} + \frac{\partial}{\partial \omega} \frac{\partial \omega}{\partial t}$$

Further we present the function $A(x, y, z, t)$ in the following form

$A(x,y,z,t) = \psi(x,y,\omega(x,y,t))\, f(x,y,z,\omega(x,y,t))$, where $f$ is the normalized eigenfunction of the problem (2.6)

$$\int_{-H(x,y)}^{0} (N^2(z) - \omega^2) f^2(x,y,z,\omega) dz = 1$$

Further we shall consider the equation (2.13) using characteristics of (2.8): in this case the function $|\mathbf{k}| = K(\omega, x, y)$ is considered as the known function. We shall multiply (2.13) by $A$ and we shall integrate with respect to the variable $z$ within the limits from $-H(x,y)$ up to zero. As a result after the rather complex transformations we shall gain

$$\psi^2 \frac{\partial^*}{\partial t}(KK'_\omega) + KK'_\omega \frac{\partial^* \psi^2}{\partial t} + 2\omega K'_\omega \psi^2 \frac{\partial^*}{\partial t}\left(\frac{K}{\omega}\right) + 2\frac{\omega}{K} \psi^2 \nabla S \nabla^*\left(\frac{K}{\omega}\right) +$$
$$+ \psi^2 \Delta S + \nabla S \nabla^* \psi^2 = 2 \nabla H \nabla S \frac{\omega^2}{K^2} (A'_z(x,y,-H(x,y),t))^2 \tag{2.15}$$

Further we shall take advantage of horizontal properties of the problem (2.6) assuming $\omega$ as a fixed spectral parameter. Further we shall apply the operator $\nabla$ to (2.6), as a result we shall gain

$$(\nabla A)''_{zz} + K^2(\omega,x,y)\left(\frac{N^2(z)}{\omega^2} - 1\right)\nabla A + A \nabla\left(K^2(\omega,x,y)\left(\frac{N^2(z)}{\omega^2} - 1\right)\right) = 0 \tag{2.16}$$

We shall multiply (2.16) by $A$ and shall integrate with respect to $z$ within the limits of from $-H(x,y)$ up to zero. Then taking (2.14) into consideration we shall have

$(A'_z(x,y,-H(x,y),t))^2 \nabla H(x,y) =$

$$= -\nabla\left(\frac{K^2}{\omega^2}\right) \int_{-H(x,y)}^{0} (N^2(z) - \omega^2) A^2(x,y,z,t) dz = -\nabla\left(\frac{K^2}{\omega^2}\right)\psi^2$$

As a result the expression (2.15) can be introduced in the following form

$$\frac{d \ln \psi^2}{dt} + \frac{d}{dt}\ln\frac{K^2}{\omega^2} - 2\frac{\nabla S \nabla \ln(K^2 \omega^{-2})}{KK'_\omega} - \frac{\partial^*}{\partial t}\ln(KK'_\omega) - \frac{\Delta S}{KK'_\omega} = 0$$

$$\frac{d}{dt} = \frac{\partial^*}{\partial t} + \frac{\nabla S}{KK'_\omega}\nabla^*$$

Here $\dfrac{d}{dt}$ is the derivative along the characteristic (2.8). As along the characteristic the frequency $\omega$ is saved, then

$$\frac{d}{dt}\ln\frac{K^2}{\omega^2} = \frac{\nabla S \nabla \ln(K^2 \omega^{-2})}{KK'_\omega}$$

From here it is possible to gain

$$\frac{d}{dt}\ln\left(\frac{\psi^2}{K^2}\right) + \frac{\Delta S}{KK'_\omega} + \frac{\partial^*}{\partial t}\ln(KK'_\omega) = 0$$

Further we have

$$\frac{\Delta S}{KK'_\omega}+\frac{\partial^*}{\partial t}\ln(KK'_\omega)=\frac{\Delta S}{KK'_\omega}+\frac{\partial^*}{\partial t}\ln(KK'_\omega)+\frac{\nabla S}{KK'_\omega}\nabla^*\ln(KK'_\omega)-\frac{\nabla S}{KK'_\omega}\nabla^*\ln(KK'_\omega)=$$

$$=\frac{d}{dt}\ln(KK'_\omega)+\frac{\Delta S}{KK'_\omega}-\frac{\nabla S}{KK'_\omega}\nabla^*\ln(KK'_\omega)=\frac{d}{dt}\ln(KK'_\omega)+\nabla^*\mathbf{c},\quad \mathbf{c}=-\frac{\nabla S}{KK'_\omega}$$

Where $\mathbf{c}$ - a vector of a group velocity of the internal waves propagation. Then keeping in mind, that along the characteristics (2.8) $\nabla^*\mathbf{c}=\operatorname{div}\mathbf{c}$, we shall finally gain

$$\frac{d}{dt}\ln\left(\frac{\psi^2 K'_\omega}{K}\right)+\operatorname{div}\mathbf{c}=0 \tag{2.17}$$

According to Liouville theorem the equation (2.17) can be written as follows

$$\frac{d}{dt}\ln(D\psi^2 K'_\omega K^{-1})=0 \tag{2.18}$$

Here D is Jacobian determinant of transition from the ray coordinates to the cartesian coordinates.

It is necessary to underline, that the gained energy conservation law (2.18), unlike the case of motion of the source in the stratified medium with the density $\rho=\rho(x,y,z)$, it is possible to consider as the conservation law of the wave energy along the ray tube. Really, from the averaged equations of hydrodynamics follows, that if the undisturbed density is the function of the horizontal coordinates, then from the existence of the stationary distribution of the density $\rho=\rho(x,y,z)$ follows existence of stationary currents However owing to the slowness of these currents they can be neglected as the first approximation. Therefore normally it is considered, that $\rho=\rho(x,y,z)$ is some background field of the density formed under action of mass forces and non-adiabatic sources and is set *a priori*, for example, from experiment.

Therefore owing to the non-equilibrium of the medium at $\rho=\rho(x,y,z)$ the energy flux is non-constant along the ray tube. However the source – the irregular bottom system considered in this paragraph is conservative, there is no energy in-leak from the outside, therefore the law (2.18) is the conservation law of the wave energy along the ray tube.

***Ray and the amplitude structures of the solution in the layer of the constant depth.***

We shall transfer to the description of the ray and amplitude structures of the solution at motion of the disturbing body in the layer of the stratified liquid of the constant depth, that later on will give the possibility to build the uniform asymptotics of a separate mode. Let the point source of mass, moving in the negative direction of axis x with the velocity V is passing the coordinates origin point at the moment $t=0$. At each moment of time the source emits the waves of all the frequencies within the range of $0<\omega<\max_z N(z)$. Frequency $\omega$ saves its value along the ray having the direction of the vector $\mathbf{k}=(p,q)$, where

$$p=\frac{\omega}{V},\quad q=(K^2(\omega)-\omega^2 V^{-2})^{\frac{1}{2}}\equiv\lambda(\omega)$$

Further it is necessary to write down the equations of the rays. As the ray coordinates it is suitable to take frequency $\omega$ and as the moment of the ray passing out of the source - $t_0$. In the case the rays ($H(x,y)=\operatorname{const}$) are the direct lines.

$$x = -Vt_0 - \frac{\omega(t-t_0)}{VK(\omega)K'(\omega)}, \quad y = \frac{\lambda(\omega)(t-t_0)}{K(\omega)K'(\omega)} \tag{2.19}$$

Using (2.19), we shall write down the expression for Jacobian $D$ and eikonal $S^*$ in the ray coordinates

$$D(t,t_0,\omega) = D(t-t_0,\omega) = V\frac{\partial^2 \lambda(\omega)}{\partial \omega^2}\left(\frac{\lambda(\omega)}{K(\omega)K'(\omega)}\right)^2 (t-t_0)$$

$$S(x,y,t) = S^*(t,t_0,\omega) = S^*(t-t_0,\omega) = \left(\omega - \frac{K(\omega)}{K'(\omega)}\right)(t-t_0) \tag{2.20}$$

It is necessary to mark, that in the coordinate system moving together with the source there is the family of rays, which is one-parametric with the argument $\omega$ and represents a fan of the rays, which are coming from the source and are arranged inside the half of the angle mouth equal to

$$\operatorname{arctg}\left(\frac{c_0}{\sqrt{V^2 - c_0^2}}\right), \quad c_0 = \frac{1}{K'(0)}.$$

The equations of rays in this coordinate system look like

$$\xi = \frac{\omega(t-t_0)}{VK(\omega)K'(\omega)}, \quad y = \frac{\lambda(\omega)(t-t_0)}{K(\omega)K'(\omega)}$$

Where $\xi = x + Vt$ is the Distance from the source up to the view point, measured along the x axis in the moving coordinate system.

Expression for the an eikonal $S^*$ in the moving coordinate system looks like

$$S^* = \frac{\omega}{V}\xi - \lambda(\omega) y \equiv \mu(\lambda)\xi - \lambda y$$

The uniform asymptotics of elevation $\eta$ at the source motion in the stratified layer of the constant depth looks like

$$\eta = Q(\xi,\lambda,z,z_0)\Phi(\sigma), \quad \sigma = \sqrt{2\xi(\mu(\lambda) - \mu'(\lambda)\lambda)}$$

$$Q(\xi,\lambda,z,z_0) = \frac{V\mu^2(\lambda)\varphi(z,\lambda)}{\sqrt{2\mu''(\lambda)\xi(\mu^2(\lambda) + \lambda^2)}} \frac{\partial \varphi(z_0,\lambda)}{\partial z_0}$$

$$\int_{-H}^{0} N^2(z)\varphi^2(z,\lambda)dz = 1, \quad \varphi^2 = f^2 \frac{K(\omega)}{\omega K'(\omega)}$$

Further, having taken advantage of the following equalities

$$\mu'(\lambda) = \frac{1}{V\lambda'(\omega)}, \quad \mu''(\lambda) = -\frac{\lambda''(\omega)}{V(\lambda'(\omega))^3}, \quad K'(\omega) = \frac{\lambda(\omega)\lambda'(\omega) + \omega V^{-2}}{K(\omega)}$$

We shalls gain expression for $\eta$ in the ray coordinates

$$\eta(t,t_0,\omega) = Q^*(t,t_0,\omega)\Phi(\sigma^*), \quad \sigma^* = \sqrt{2S^*(t,t_0,\omega)} \tag{2.21}$$

$$Q^*(t,t_0,\omega) = \frac{\omega f(z,\omega)\sqrt{K(\omega)K'(\omega)}}{V\lambda(\omega)\sqrt{2|\lambda''(\omega)|\lambda(\omega)(t-t_0)}} \frac{\partial f(z_0,\omega)}{\partial z_0}$$

From here we have the obvious expression for the amplitude $\psi$ for the case of the layer of the constant depthl

$$\psi = \frac{\omega\sqrt{K(\omega)K'(\omega)}}{V\lambda(\omega)\sqrt{2|\lambda''(\omega)|\lambda(\omega)(t-t_0)}} \frac{\partial f(z_0,\omega)}{\partial z_0} \tag{2.22}$$

***Uniform asymptotics of a separate mode.***

The energy conservation law (2.18) can be rewritten in the ray coordinates $l, t_0$ in the following form

$$D(t,t_0,l)\psi^2(t,t_0,l)\frac{K'_\omega(t,t_0,l)}{K(t,t_0,l)} = E(t_0,l) \tag{2.23}$$

Where $E(t_0,l)$ is a function being determined by the particular kind of the being solved problem and which one can find using the solution of the problem concerning the motion of the point source of mass in the layer of the stratified liquid of the constant depth. For this purpose we shall take advantage of the principle of locality, that is we shall consider, that on the typical distances, on which the uniform asymptotics (2.21) in the layer of the constant depth (the order of several depths of the layer) is true, the depth of the a bottom may be considered as locally-constant. Thus, for these distances the layer of the stratified liquid has locally-constant depth. In this case the function standing in the right part of (2.23), may be calculated with assumption of the local constant depth with the "frozen" horizontal parameters.

Assume, that at the moment $t = t_0$ the moving source is in the point $x_0, y_0, z_0$. Then, using (2.20), (2.22), it is possible to gain

$$E(t_0,\omega) = \frac{\omega^2}{2V\lambda(\omega,x_0,y_0)K^2(\omega,x_0,y_0)}\left(\frac{\partial f(x_0,y_0,z_0,\omega)}{\partial z_0}\right)^2$$

As a result the first term of the uniform asymptotics of elevation of Fresnel wave at motion of the source in the later of the stratified liquid with the slowly-varying bottom will look like

$$\eta = \eta_0\Phi(\sigma^*)f(x,y,z,\omega), \quad \sigma^* = \sqrt{2S(t,t_0,\omega)}$$

$$\eta_0(t,t_0,\omega) = \frac{\partial f(x_0,y_0,z_0,\omega)}{\partial z_0} \frac{\omega}{\sqrt{2V\lambda(\omega,x_0,y_0)K^2(\omega,x_0,y_0)}} \sqrt{\frac{K(\omega,x,y)}{K'_\omega(\omega,x,y)D(t,t_0,\omega)}}$$

$$x = x(t,t_0,\omega), \quad y = y(t,t_0,\omega)$$

Thus, the solution gained in the present paragraph in the most general view describes the uniform asymptotics of the far field of the internal gravity waves at motion of the source above the slowly-varying bottom since at $H(x,y) = \text{const}$ this solution coincides with the uniform asymptotics. At the great values of $\sigma^*$ (far from the wave front), having taken advantage of the asymptotical behavior of Fresnel integrals at the great values of the argument, it is possible to gain the notation of the solution in the form of the normal WKB (Wentzel-Kramers-Brillouin) expansion with respect to the locally-harmonious waves. At the small values of $\sigma^*$ the gained solutions describe the asymptotics of the field in the vicinity of the wave front of the separate mode.

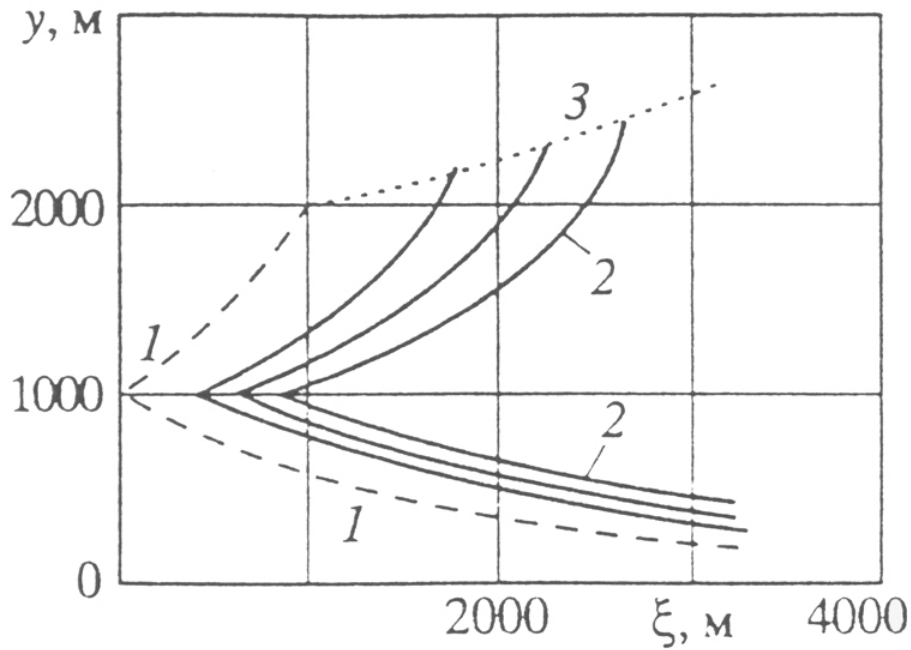

Fig.1

## 3. Local asymptotics of the distant field of the internal gravity waves in the layer of the stratified medium with smoothly varying bottom

In the present section, we consider one of the versions of the "traveling wave" method, which is one of the modifications of the geometrical optics method, that is the problem about the distant field of the internal gravity waves in the vicinity of the separate mode front, generated by the point source of mass moving above the smoothly varying bottom. The problem is considered in the so-called weakly disperse approximation, that is the gained solution describes the wave field only near to the corresponding wave front (the local asymptotics).

***Formulation of the problem and selection of the way of its the solution.***

We shall consider the liquid layer with Brunt-Väisälä frequency $N(z)$, limited by the surface $z = 0$ and in the bottom $z = H(X, Y)$. The point source of the unity intensity moves homogeneously and rectilinearly with the velocity V on the depth $z_0$ in the positive direction of the abscissa axis. Then the field of the velocities in Boissinesq approximation meets the conditions of the following linearized system of equations

$$\frac{\partial^2}{\partial T^2}\left(\Delta w + \frac{\partial^2 w}{\partial z^2}\right) + N^2(z)\Delta w = \delta''_{TT}(X - VT)\delta(Y)\delta'(z - z_0), \tag{3.1}$$

$$\Delta \mathbf{u} + \nabla \frac{\partial w}{\partial z} = \delta(z - z_0)\nabla(\delta(X - VT)\delta(Y)).$$

Here $\nabla = \left(\frac{\partial}{\partial X}, \frac{\partial}{\partial Y}\right)$; $\Delta = \frac{\partial^2}{\partial X^2} + \frac{\partial^2}{\partial Y^2}$; $w$ is the vertical component of velocity; $\mathbf{u} = (u_1, u_2)$ - the vector of horizontal velocities. It is supposed, that on the borders of the layer there are no leakage conditions

$$w = 0 \text{ At } z = 0, \tag{3.2}$$
$$w = \mathbf{u}\cdot\nabla H(X,Y) \text{ At } z = H(X, Y).$$

Introduce the dimensionless parameter $\varepsilon = \lambda / L \ll 1$, describing the smoothness of variation of the bottom depth, $\lambda$ - the reference length of the wave of the internal waves, L - the horizontal scale of variation of the bottom depth. Then in the "slow variables" $x = \varepsilon X, y = \varepsilon Y, t = \varepsilon T$ (on the vertical coordinate $z$ the slowness of variation is not expected) the equations of the motion (3.3.1) and the boundary conditions (3.2) may be written as

$$\frac{\partial^2}{\partial t^2}\left(\varepsilon^2 \Delta w + \frac{\partial^2 w}{\partial z^2}\right) + N^2(z)\Delta w = \varepsilon^2 \delta''_{tt}(x - vt)\delta(y)\delta'(z - z_0), \tag{3.3}$$

$$\varepsilon \Delta \mathbf{u} + \nabla \frac{\partial w}{\partial z} = \varepsilon^2 \delta(z - z_0)\nabla(\delta(x - Vt)\delta(y));$$

$$w = 0 \text{ при } z = 0, \quad w = \varepsilon \mathbf{u}\cdot\nabla h(x, y) \text{ при } z = h(x, y). \tag{3.4}$$

$$F_n(0, \omega) = F_n(h, \omega)$$

where the function h (x, y) describes the shape of the bottom in the "slow" variables.
The solution for the case of the constant depth h of the layer for w as the sum of the modes $w = \sum_{n=1}^{\infty} w_n$. In the same place there are the first terms of the asymptotics for $w_n$ near to the front and expressed through Airy function derivative or Fresnel integrals, which argument depends on the first two coefficients of the corresponding expansions of the dispersion curve $k_n(\omega) = c_n^{-1}\omega + d_n\omega^3 + \Lambda$ in zero (Airy wave), where $k_n(\omega)$ is the eigenvalue of the spectral problem

$$\frac{\partial^2 F_n(z,\omega)}{\partial z^2} + k_n^2(\omega)\left[\frac{N^2(z)}{\omega^2} - 1\right] F_n(z,\omega) = 0,$$

$$F_n(0,\omega) = F_n(h,\omega) \tag{3.5}$$

Further for distinctness and without limitation of generality we shall consider Airy wave propagation. The solution of the system (3.3), (3.4) we shall also search in the form of the sum of modes

$$w = \sum_{n=1}^{\infty} w_n,$$

$$\mathbf{u} = \sum_{n=1}^{\infty} \mathbf{u}_n.$$

Later on all calculations we shall refer to the separately taken mode, omitting the index $n$. Proceeding from the aforesaid, and also from the structure of the asymptotics of the solution in the layer of the constant depth, the solution of the system (3.3), (3.4), is similar to one given in the section 2, we shall search it in the form of

$$w = \varepsilon^{(2/3)} \sum_{i=0}^{\infty} \sum_{k=0}^{\infty} \varepsilon^{(2/3)(k+i)} w_{ik} R_i(\varphi), \tag{3..6}$$

$$\mathbf{u} = \varepsilon^{(1/3)} \sum_{i=0}^{\infty} \sum_{k=0}^{\infty} \varepsilon^{(2/3)(k+i)} \mathbf{u}_{ik} R_{i+1}(\varphi)$$

($\frac{\partial R_i(\varphi)}{\partial \varphi} = R_{i-1}(\varphi)$, $R_0(\varphi) = Ai'(\varphi)$ is Airy function derivative,

$\varphi = \varepsilon^{-2/3}((t - \tau(x,y))\sigma(x,y))$, at that the argument $\varphi$ we assume as the unity order).

As further we shall be interested only in the first term of an asymptotics for $w$, then we shall to write (3.6) out in the following form

$$w = \varepsilon^{2/3} A(z,x,y) v_0(\varphi) + \varepsilon^{4/3}(B(z,x,y) v_0(\varphi) + C(z,x,y) v_1(\varphi)) + O(\varepsilon^2),$$

$$\mathbf{u} = \varepsilon^{1/3} \mathbf{u}_0(z,x,y) v_1(\varphi) + O(\varepsilon). \tag{3.7}$$

Functions $A(z,x,y)$, $\mathbf{u}_0(z,x,y)$, $\tau(x,y)$, $\sigma(x,y)$ are the subject to determination. Substituting (3.7) in the second equation (3.3) and equating terms at zero degree of the small argument $\varepsilon$, we shall have $\mathbf{u}_0 = A'_z(z,x,y)\nabla \tau(x,y)/(\sigma(x,y)|\nabla \tau(x,y)|^2)$.

Boundary conditions for $A, B, C$ functions we shall find substituting (3.7) in (3.4)

$A = B = C = 0$ at $z = 0$,

$A = B = 0$, $C = A'_z \nabla \tau \nabla h /(\sigma|\nabla \tau|^2)$ при $z = h(x,y)$.

*Derivation of the fundamental equations.*

Now we shall begin derivation of the equations for the functions $\tau(x,y)$, $A(z,x,y)$ and $\sigma(x,y)$. Substituting (3.7) in the first equation (3.3) and equating terms of the order $\varepsilon^{2/3}$, we shall gain

$$\frac{\partial^2 A(z,x,y)}{\partial z^2} + |\nabla \tau(x,y)|^2 N^2(z) A(z,x,y) = 0 \tag{3.8}$$

$A(0,x,y) = A(h(x,y),x,y) = 0.$

We shall note, that eigenfunctions $A(z, x, y)$ are determined from (3.8) with accuracy up to the arbitrary multiplier depending only on the horizontal $x$ and $y$, therefore it is convenient to present the function $A(z, x, y)$ in the form of $A(z, x, y) = \psi(x, y) f(z, x, y)$, where $f(z, x, y)$ is the solution of the spectral problem (3.8) and meets the condition of normalization

$$\int_0^{h(x,y)} N^2(z) f^2(z, x, y) \, dz = 1. \qquad (3.9)$$

Eigenfunctions $f(z, x, y)$ and the eigennumbers $\lambda(x, y)$ of the problem (3.8) are supposed to be known. Then for $\tau(x, y)$ we have the eikonal equation

$$\left(\frac{\partial \tau}{\partial x}\right)^2 + \left(\frac{\partial \tau}{\partial y}\right)^2 = \lambda^2(x, y). \qquad (3.10)$$

To find the functions $\psi(x, y)$ and $\sigma(x, y)$ we should after substitution (3.7) in (3.3) to equate the terms of the order $\varepsilon^0$. Having taken advantage of the properties of Airy function $R_4(\varphi) = -\varphi R_2(\varphi) - 3 R_1(\varphi)$ we shall gain two equations. The first equation- with $B$, is containing the terms with phase function $R_2$,

$$\sigma^2 (B''_{zz} + \lambda^2 N^2(z) B) = 2 \varphi A N^2(z) \nabla \sigma \nabla \tau + \varphi A \sigma^4 \lambda^2, \qquad (3.11)$$

$B = 0$ at $z = 0, h(x, y)$;

and the second equation - with $C$ is containing the terms with the phase function $R_1$

$$\sigma^2 (C''_{zz} + \lambda^2 N^2(z) C) = 2 \sigma N^2(z) \nabla A \nabla \tau + A N^2(z) (2 \nabla \sigma \nabla \tau + \sigma \Delta \tau) + 3 A \sigma^4 \lambda^2, \qquad (3.12)$$

$C = 0$ at $z = 0$, $C = A'_z \nabla \tau \nabla h /(\sigma \lambda^2)$ at $z = h(x, y)$.

First we shall consider the equation (3.11) all over again. Multiplying its both parts by function $A(z, x, y)$ and integrating it by $z$ from 0 up to $h(x, y)$, we shall gain the equation for $\sigma$

$$2 \nabla \sigma \nabla \tau + a(x, y) \lambda^2 \sigma^4 = 0$$

$$\left( a(x, y) = \int_0^{h(x,y)} f^2(z, x, y) \, dz \right). \qquad (3.13)$$

Where $c(x, y)$, as before is the group velocity at $\omega = 0$:

$$c(x, y) = \left[ \frac{\partial k(\omega, x, y)}{\partial \omega} \right]_{\omega=0}^{-1}.$$

Further we shall consider the equation (3.12) and multiply its both parts by $A(z, x, y)$ and integrate by $z$ from 0 up to $h(x, y)$. Taking consideration of the conditions of normalization (3.9), we shall gain

$$-\sigma\lambda^{-2}\psi^2[f'_z(h, x, y)]^2 \nabla\tau\nabla h = \sigma\ \nabla\tau\nabla\psi^2 + \psi(2\nabla\tau\nabla\sigma + \sigma\Delta\tau) + 3\psi^2\sigma^4\lambda^2 a. \quad (3.14)$$

Differentiating the equation (3.8) by the horizontal variables it is easy to demonstrate, that $[f'_z(h, x, y)]^2 \nabla h(x, y) = -\nabla\lambda^2(x, y)$.

Then the transportation equation (3.14) may be presented in the following form

$$\nabla \ln\left(\frac{\psi^2}{\lambda^2\sigma^4}\right)\nabla\tau + \Delta\tau = 0. \quad (3.15)$$

So the plotting of the field $w$ in the form of the expression in (3.7) was reduced to the solution of the eikonal equation (3.10) and the transportation equations (3.13) и (3.15).

*The solution of the equations of the eikonal and transportation.*

The characteristic system for (3.10) looks like this $(p = \partial\tau/\partial x, \quad q = \partial\tau/\partial y)$:

$$\frac{dx}{d\tau} = c^2(x, y)p, \quad \frac{dy}{d\tau} = c^2(x, y)q,$$

$$\frac{dp}{d\tau} = -c'_x/c(x, y), \quad \frac{dq}{d\tau} = -c'_y/c(x, y). \quad (3.16)$$

So from the selection of this characteristic system is clear, that $\frac{\partial\tau}{\partial\tau} = 1$, therefore in the capacity of the parameter of integration it is convenient to take $\tau$ eikonal. The solution of the system (3.16) is the one-parameter family of functions $x(\tau, \tau_0)$, $y(\tau, \tau_0)$, $p(\tau, \tau_0)$, $q(\tau, \tau_0)$, the first two functions of which determine $x, y$ ray on the plane, $\tau_0$ is the initial eikonal, or, that is the same, the time of the ray outlet from the source. Let's suppose, that the source is moving along the axis $y = 0$ and passes through the point of origin of coordinates during the moment $\tau = 0$. Then we have the initial conditions for system (3.16):

$$x_0 = V\tau_0, \quad y_0 = 0, \quad p_0 = \frac{1}{V}, \quad q_0 = \pm\sqrt{\frac{1}{c^2(x_0,0)} - \frac{1}{V^2}}. \tag{3.17}$$

The equations of the rays $x = x(\tau, \tau_0)$, $y = y(\tau, \tau_0)$ at the fixed $\tau_0$ determine the particular ray and at the fixed $\tau$ - determine the wave front. We shall suppose, that the equations of rays are solvable with respect to $\tau$ and $\tau_0$:

$$\tau = \tau(x, y), \quad \tau_0 = \tau_0(x, y). \tag{3.18}$$

For this purpose it is necessary, that the Jacobian $D \equiv \frac{\partial x}{\partial \tau}\frac{\partial y}{\partial \tau_0} - \frac{\partial y}{\partial \tau}\frac{\partial x}{\partial \tau_0}$ was not = 0.

Thus, the equations (3.18) for the point $x, y$ determine the eikonal $\tau$ (the moment of the wave front coming in the point $x, y$) and the initial eikonal $\tau_0$, (the moment of the ray leaving the source).

Transportation equations (3.13) and (3.15) are integrated along the characteristics (3.16), and the applicable quadrature for (3.13) looks like

$$\sigma(x, y) = \left[ \frac{3}{2} \int_{\tau_0(x,y)}^{\tau(x,y)} a(x(t, \tau_0), y(t, \tau_0))\, dt \right]^{-1/3}. \tag{3.19}$$

In view of expression along the ray: $\Delta\tau = \nabla \ln\left(\frac{J}{c}\right)\nabla\tau$, where $J(x, y)$ is the geometrical ray divergence of the ray tube is connected with the Jacobian by the known ratio $J = D/c$, integration of the equation (3.15) gives some "conservation law" along the ray in the form of

$$c(x, y)\psi^2(x, y)J(x, y)/(\sigma^4(x, y)J(x_0, 0)) = B(x_0).$$

Here $J(x, y)$ and $J(x_0, 0)$ - the geometrical ray divergence of the rayl tube at the front and in the point of the ray outlet accordingly, $J(x_0, 0) = \sqrt{V^2 - c^2(x_0, 0)}$. $B(x_0)$ constant similarly to the section 2 can be determined from the solution of the problem with the constant depth of the bottom $h(x_0, 0)$.

$$B(x_0) = c^3(x_0,0) f'_z(z_0,x_0,0) / \left[4(V^2 - c^2(x_0,0))\right].$$

Let's write down the resultant expression

$$\psi(x,y) = \frac{\sigma^2(x,y)(V^2 - c^2(x_0,0))^{1/2} c^{3/2}(x_0,0) f'_z(z_0,x_0,0)}{2 c^{1/2}(x,y) J^{1/2}(x,y)}. \qquad (3.20)$$

Thus, we have the following diagram for the search of the field of the vertical velocity in the vicinity of the wave front of the moving source:

a) we solve the characteristic system (3.16) with the initial conditions (3.17);

b) having solved the equations of the rays, we shall find the eikonal $\tau(x,y)$ and the moment, when the ray leaves the source $\tau_0(x,y)$;

c) solving the boundary-value problem (3.8) we shall gain the normalized eigenfunction - $f(z,x,y)$ and the coefficient $a(x,y)$;

d) integrating $a(x,y)$ along the ray, we determine $\sigma(x,y)$ from (3.19);

e) we search for the geometrical ray divergence $J$, for example, using numerical differentiation;

f) calculating the function $\psi(x,y)$ with the help of (3.20) and multiplying it by $f(z,x,y)$, we shall have amplitude $A(z,x,y)$;

g) multiplying the amplitude $A(z,x,y)$ by the derivative of Airy function of argument $\varphi$, we shall gain the vertical velocity of the separately taken mode.

Further, without loss of generality, we shall dwell on the case, when Brunt-Väisälä frequency $N = \text{const}$, and depth of the bottom depends linearly only on one coordinate $H(y) = \beta y$. Now we shall introduce the coordinate system with the axis $x$, going along the "shore" $(y = 0)$, the source is radiant moving from left to right in the positive direction of the axis $x$ with rate $V$ in parallel to the "shore" at the distant $y_0$ from it{him} and on the depth $z_0$.

Let's consider the first mode. Then the equation (3.8) gives the following eigenfunction $f(z,y)$ and the eigenvalue $\lambda(y)$ ($\gamma = N\beta/\pi$):

$$f(z,y) = \frac{\sqrt{2}}{N\sqrt{\beta y}} \sin \frac{\pi z}{\beta y}, \quad \lambda(y) \equiv \frac{1}{c(y)} = \frac{1}{\gamma y}. \qquad (3.3.21)$$

We shall write down the characteristic system and the initial conditions for the equation of an eikonal

$$\dot{x} = \gamma^2 y^2 / V, \quad x_0 = V\tau_0, \quad \dot{y} = \pm \gamma y \sqrt{1 - (\gamma y / V)^2}, \quad y_0 = y_0. \qquad (3.22)$$

Here and further the upper character corresponds to the area $y > y_0$, and the lower character - to the area $y < y_0$.

Integrating the system (3.22), we shall gain the equations of the rays

$$y = \frac{V}{\gamma} ch^{-1}\left(\pm \operatorname{arch}\left(\frac{V}{\gamma y_0}\right) - \gamma(\tau - \tau_0)\right), \quad x = x_0 + \frac{\gamma}{V} y_0 y \, sh(\gamma(\tau - \tau_0)) \qquad (3.23)$$

( $\operatorname{arch} x = \ln(x + \sqrt{x^2 - 1})$ )

The rays set by the system (3.23) are semicircles of radius $V/\gamma$ with the centers arranged along the shore", and at that these semicircles at $y = V/\gamma$ have the envelope (caustic) and in the further we shall consider the field outside the vicinity of the caustic and the "shore".

As in this case the pattern of the waves is stationary in the moving together with the source coordinate system ($\xi = Vt - x$), then the front is determined from the equation

$$\frac{d\xi}{dy} = \frac{\pm \sqrt{V^2 - (\gamma y)^2}}{\gamma y}, \quad \xi(y_0) = 0 \qquad (3.24)$$

and looks like

$$\xi = \pm \frac{V}{\gamma}(\alpha_1(y) - \alpha_2(y)), \qquad (3.25)$$

$$\alpha_1(y) = \operatorname{arch}\left(\frac{V}{\gamma y_0}\right) - \operatorname{arch}\left(\frac{V}{\gamma y}\right), \quad \alpha_2(y) = \sqrt{1 - \left(\frac{\gamma y_0}{V}\right)^2} - \sqrt{1 - \left(\frac{\gamma y}{V}\right)^2}.$$

the equations of the rays (3.23) are solved with respect to $\tau$ и $\tau_0$:

$$\tau = \frac{x}{V} \pm \frac{1}{\gamma}(\alpha_1(y) - \alpha_2(y)), \quad \tau_0 = \frac{x}{V} \mu \frac{1}{\gamma}\alpha_2(y).$$

Coefficient $a(x, y) = N^{-2}$, from here

$$\sigma(y) = \frac{1}{\left(\pm\frac{3}{2} N^{-2}\gamma^{-1}\alpha_1(y)\right)^{1/3}}. \qquad (3.26)$$

Now we shall write down the expression for the argument $\varphi(\xi, y)$ of the derivative of Airy function:

$$\varphi(\xi, y) = \frac{\left(\frac{\xi}{V}\mu\frac{(\alpha_1(y) - \alpha_2(y))}{\gamma}\right)}{\left(\pm\frac{3}{2}\frac{\alpha_1(y)}{N^2\gamma}\right)^{1/3}}. \qquad (3.27)$$

Having taken advantage of Liouville theorem, we shall gain the geometrical ray divergence $J = \sqrt{V^2 - \gamma^2 y^2}$. Thus, all the components taking part in the solution for w are discovered and the final expression looks like the following

$$w = \frac{\sigma^2(y) c^{3/2}(y_0)}{2c^{1/2}(y)}\left(\frac{V^2 - c^2(y_0)}{V^2 - c^2(y)}\right)^{1/4} f_z'(z_0, y_0) f(z, y) \, Ai'(\varphi(\xi, y)) \qquad (3.28)$$

(Functions $c(y)$, $f(z,y)$, $\sigma(y)$ and $\varphi(\xi, y)$ are determined from (3.21), (3.26) and (3.27) accordingly).

The Figures 2 and 3 present the results of the numerical calculations in the dimensionless variables $\xi^* = \xi\gamma/V$, $y^* = y\gamma/V$, $z^* = z/\beta y_0$, $Q^* = QN^2/V^3$, $w^* = w/V$.

On the Fig. 2 there are presented the left-hand and the right-hand fronts calculated by the formula (3.25) at $y_0^* = 0,4$.

On the Fig. 3 there are the continuous lines - the diagrams of the vertical velocity $w^*(\xi^*)$ built by the formula (3.28) for the values of $Q^* = 1$, $z_0^* = 0,2$, $z^* = 0,1$ and $y^* = 0,29$(a), $y^* = 0,51$(b); the dashed lines - the vertical velocity at the constant depth $H^* = 1$. From the presented numerical results one can see, for example, that to the left of the axis of the motion the wave amplitude for the variable bottom is less, than for the constant bottom, and to the right the wave amplitude is higher.

Thus, the results gained in this section describe the local (near to the corresponding wave front) asymptotics of the far field of the internal gravity waves oscillated by the moving sources of disturbance in the layer of the stratified medium with the slowly variable bottom.

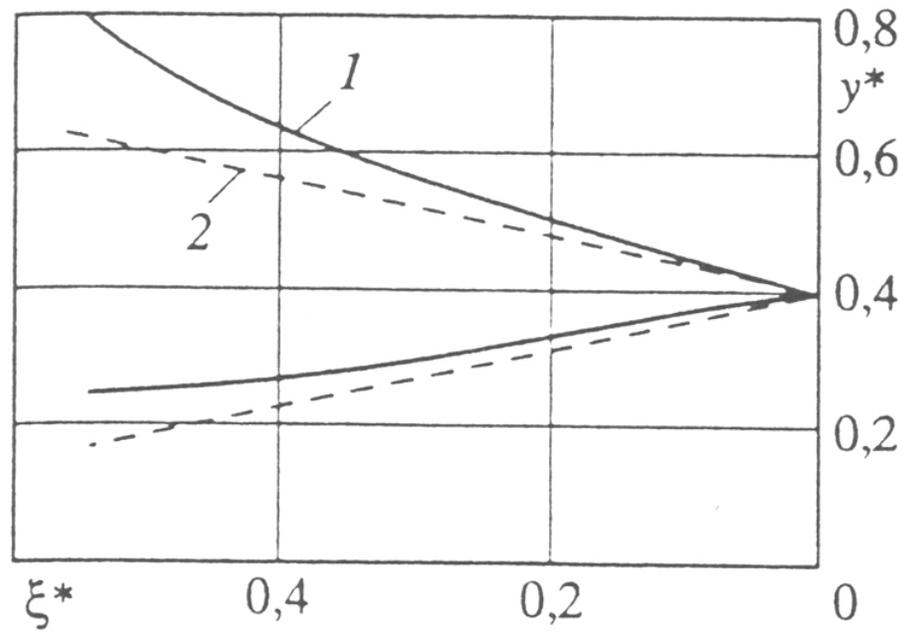

Fig.2

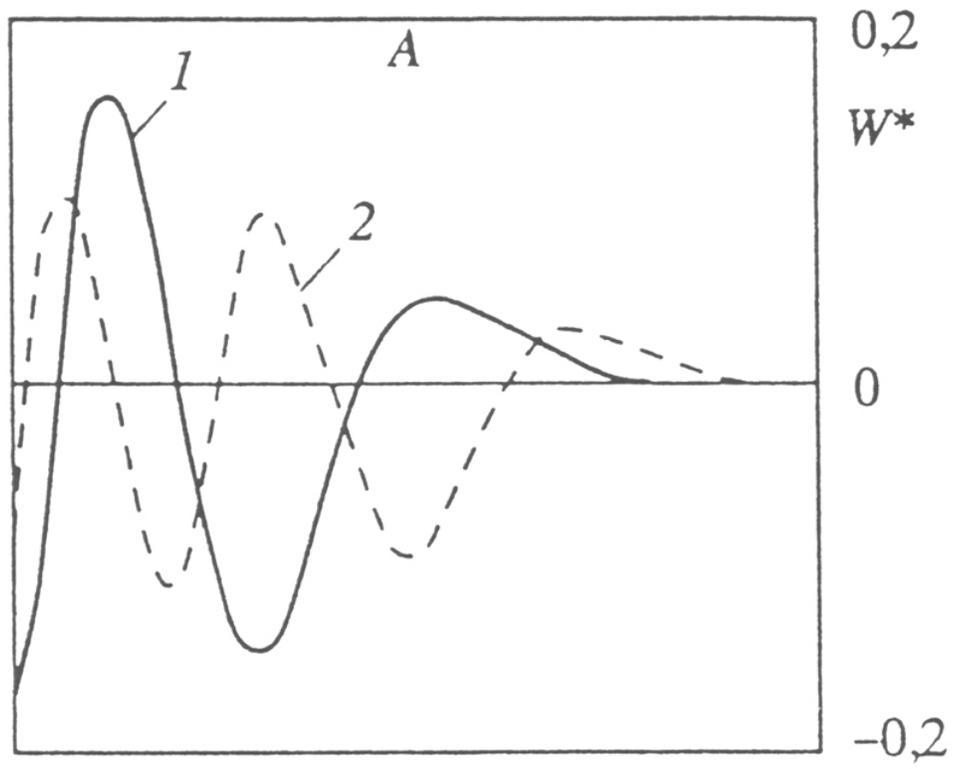

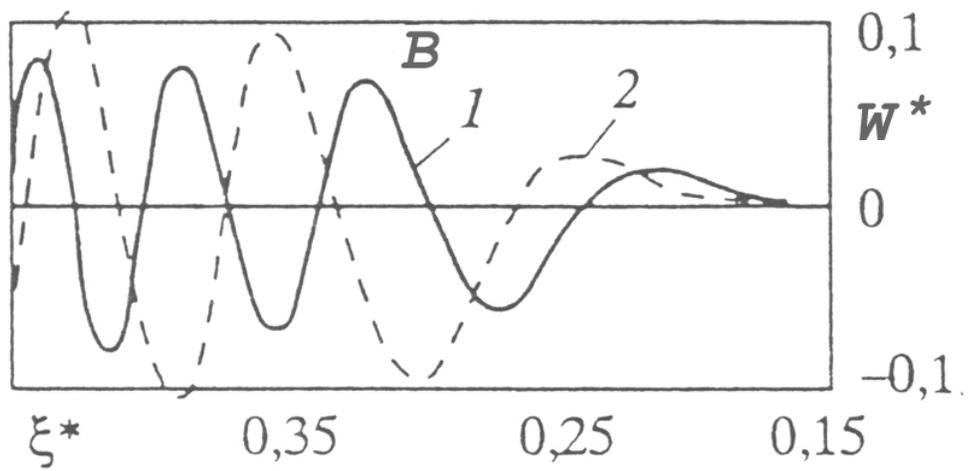

Fig.3

## 4. Uniform asymptotics of the far field of the internal gravity waves in the stratified heterogeneous in the horizontal direction mediums

As the main parameter defining the behavior of the internal gravity waves is Brunt-Väisälä frequency $N^2 = -g\,\partial \ln\rho/\partial z$ ( g is acceleration of gravity, $\rho$ is the undisturbed density), and, as it is obvious, that in the real conditions the density $\rho$ depends not only on the vertical variable z, but also on the horizontal variables x, y.

Therefore it is of interest to us to consider the problem view a problem of the internal waves generated by the different sources, in particular, by the source of mass in the medium with the varying at all variables density $\rho$ and $N^2$.accordingly. Difficulty of such a problem is stipulated by the fact, that at $\rho = \rho(z, x, y)$ the equations bearing the partial derivatives describing the internal gravity waves exclude separations of the variables. However the feature, that the characteristic horizontal scale of the variation of the density is great, as compared with the characteristic lengths of the internal waves, enables to apply to the solution of this problem the method of approximation analogous to the method of the ray optics which has been already descibed in the previous sections.

We shall consider the problem of construction of the homogeneous asymptotics of the distant field of the internal gravity waves in the stratified heterogeneous in the horizontal direction medium, that is the problem of propagation of the global Airy and Fresnel internal waves generated by a point source of mass moving with the velocity V along the axis x on the depth $z_0$ in the stratum $-h < z < 0$ of the stratified horizontally heterogeneous liquid with the density $\rho = \rho(z, \varepsilon x, \varepsilon y)$ (the small argument $\varepsilon$, as usually, characterizes the "slow variables"; the slowness of variation of the variable z is not expected). We shall consider, that in the medium with the heterogeneous in the horizontal direction field of density, it is possible to neglect the steady flows caused by this field. Then from the linearized system of equations of the hydrodynamics in Boissinesq approximation it is possible to gain

$$\frac{\partial^2}{\partial t^2}\left(\Delta + \frac{\partial^2}{\partial z^2}\right)w - \frac{g}{\rho}\Delta(\mathbf{v}\,grad\,\rho) = \delta''_{tt}(x-Vt)\,\delta(y)\,\delta'(z-z_0) \qquad (4.1)$$

$$\Delta\mathbf{u} + \nabla\frac{\partial w}{\partial z} = \delta(z-z_0)\,\nabla[\delta(x-Vt)\,\delta(y)] \qquad (4.2)$$

$$\nabla = \left(\frac{\partial}{\partial x}, \frac{\partial}{\partial y}\right), \qquad \mathbf{v} = (\mathbf{u},\ w), \qquad \Delta = \frac{\partial^2}{\partial x^2} + \frac{\partial^2}{\partial y^2}$$

where w is the vertical component of the internal waves velocity, $\mathbf{u}$ is the vector of the horizontal velocities. As the boundary conditions we take: $w = 0$ at $z = 0, -h$. The solution described in (4.1), (4.2) is similar to the solution from sections 2 - 3, but we shall search in the form of the sum of modes; in the further all the calculations we shall refer to the separately taken mode neglecting its index. Proceeding from the structure of the global asymptotics for the horizontally-homogeneous medium, the solution of (4.1) - (4.2) we shall search in the following form

$$w = A(\varepsilon x, \varepsilon y, z, \varepsilon t)F_0(\sigma) + \varepsilon^a B(\varepsilon x, \varepsilon y, z, \varepsilon t)F_1(\sigma) + O(\varepsilon^{2a}) \qquad (4.3)$$

$$\mathbf{u} = \mathbf{u}_0(\varepsilon x, \varepsilon y, z, \varepsilon t)\,F_1(\sigma)\,\varepsilon^{a-1} + O(\varepsilon^{2a-1}) \qquad (4.4)$$

$$F'_{m+1}(\sigma) = F_m(\sigma), \qquad \sigma \equiv \left(\frac{S(\varepsilon x, \varepsilon y, \varepsilon t)}{a\varepsilon}\right)^a$$

Where the argument $\sigma$ is considered of the order of unity; the functions $S$, $A$, $\mathbf{u}_0$ are the subject to determination. The value of $a$ for Airy wave is equal to $2/3$, for Fresnel wave $a = 1/2$. If to search for the solution for the vertical component of velocity of Airy wave, then in the capacity of $F_0(\sigma)$ it is necessary to take the derivative of Airy function $Ai'(\sigma)$, for elevation of Airy wave in the capacity of $F_0(\sigma)$ it is necessary to take Airy function $Ai(\sigma)$. For the vertical component of the velocity of Fresnel wave in the capacity of $F_0(\sigma)$ we shall take $\Phi'(\sigma)$ and accordingly - for elevation of Fresnel wave

$$F_0(\sigma) = \Phi(\sigma), \quad \Phi(\sigma) = \mathrm{Re}\int_0^\infty \exp\left(-it\sigma - i\frac{t^2}{2}\right)dt$$

$$\mathbf{u}_0 = -A'_z \left(\frac{S}{a}\right)^{1-a} \frac{\nabla S}{\varepsilon\left(\left(\frac{\partial S}{\partial x}\right)^2 + \left(\frac{\partial S}{\partial y}\right)^2\right)} \tag{4.5}$$

We shall substitute (4.3) - (4.5) in the equation (4.1) and having equated terms at equal powers of $\varepsilon$, we gain at $\varepsilon^a$

$$\frac{\partial^2 A}{\partial z^2} + k^2\left(\frac{N^2(z,x,y)}{\omega^2} - 1\right)A = 0 \tag{4.6}$$

$$A = 0, \quad z = 0, -h$$

Where the following characters are introduced: $\mathbf{k} \equiv (p,q) = \nabla S$, $\omega = \partial S/\partial t$.
At the solution of the gained problem the dispersion dependence designated through $K(\omega, x, y)$, is supposed to be known, then for determination of $S$ function we have the eikonal equation

$$\left(\frac{\partial S}{\partial x}\right)^2 + \left(\frac{\partial S}{\partial y}\right)^2 = K^2(\omega, x, y) \tag{4.7}$$

The equation (4.7) is Hamilton-Jacobi equation with Hamiltonian function $H(\omega, \mathbf{k}, x, y) \equiv |\mathbf{k}|^2 - K^2(\omega, x, y)$. The characteristic system for the eikonal equation (4.7) looks like

$$\frac{dx}{d\tau} = \frac{p}{K(\omega,x,y)K'_\omega(\omega,x,y)}, \quad \frac{dy}{d\tau} = \frac{q}{K(\omega,x,y)K'_\omega(\omega,x,y)}$$

$$\frac{dp}{d\tau} = \frac{K'_x(\omega,x,y)}{K'_\omega(\omega,x,y)}, \quad \frac{dq}{d\tau} = \frac{K'_y(\omega,x,y)}{K'_\omega(\omega,x,y)}, \quad \frac{d\omega}{d\tau} = 0 \tag{4.8}$$

The initial conditions for the system (4.8) it is convenient to set in the three-dimensional space $x, y, t$ on some surface: $t = t_0$, $x = x_0(l)$, $y = y_0(l)$. Let the eikonal $S$ is known on this surface: $S(x, y, t) = S_0(l, t_0)$. This conforms to the assignment of the initial eikonal onto some stationary line $x = x_0(l)$, $y = y_0(l)$ in the arbitrary moment of time $t_0$. Differentiating the initial eikonal by $l, t_0$, we shall gain the system of equations for determination of the initial values of frequency $\omega_0$ and the wave vector

$$(p_0, q_0) \qquad \omega_0(l, t_0) = \frac{\partial S_0}{\partial t_0}$$

$$p_0(l, t_0) x_0'(l) + q_0(l, t_0) y_0'(l) = \frac{\partial S_0}{\partial t_0}$$

$$p_0^2(l, t_0) + q_0^2(l, t_0) = K^2(\omega_0(l, t_0), x_0(l), y_0(l))$$

Thus, the solution of the system (4.8) determines in the three-dimensional space $x, y, t$ the set of the space-time rays $x = x(t, t_0, l)$, $y = y(t, t_0, l)$, where $x = x_0(l)$, $y = y_0(l)$ at $t = t_0$. At that $l$ and the moment of the ray $t_0$ leaving the source are acting in the role of the ray coordinates, and the variable $t$ is simultaneously Cartesian and the ray coordinate. Projections of the space-time rays onto the plane $x, y$ determines unlike the characteristic system for a monochromatic wave the two-parameter set of the rays, which at fixed $t_0$ transforms into the normal one-parameter set of the rays with ray coordinates $l$ and $t$ and consequently the variables $l$ and $t_0$ in the further we shall call the ray coordinates. The frequency $\omega$, as one can see from (4.8), saves along the ray its initial value $\omega_0(l, t_0)$ and the eikonal $S^*(t, t_0, l) = S(x(t, t_0, l), y(t, t_0, l), t)$ in the ray coordinates is determined by its integration along the ray

$$S^*(t, t_0, l) = S_0(l, t_0) + \omega_0(l, t_0)(t - t_0) + \int_{t_0}^{t} \frac{K(\omega_0(l, t_0), x(\tau, t_0, t), y(\tau, t_0, t))}{K_\omega'(\omega_0(l, t_0), x(\tau, t_0, t), y(\tau, t_0, t))} d\tau$$

To find $S(x, y, t)$ and $\omega(x, y, t)$ in the cartesian coordinates, it is enough to convert the rays equations $l = l(t, x, y)$, $t_0 = t_0(t, x, y)$ and for this purpose it is necessary, that at any $t$ the Jacobian

$$D(t, t_0, l) \equiv \frac{\partial x}{\partial t_0} \frac{\partial y}{\partial l} - \frac{\partial x}{\partial l} \frac{\partial y}{\partial t_0}' \quad \text{was not equal to zero.}$$

To determine the amplitude dependence of $A$, we shall preliminary introduce $A(x, y, z, t)$ in the form of $A(x, y, z, t) = A^*(x, y, z, \omega(x, y, t)) = \psi(x, y, \omega) f(x, y, z, \omega)$, where $f$ is the normalized eigenfunction of the problem (4.6)

$$\int_{-h}^{0}[N^2(z,x,y)-\omega^2]f^2(x,y,z,\omega)dz=1$$

Then substituting (4.3) - (4.5) in (4.1) and using the properties of Airy function and Fresnel integrals, equating the terms at $\varepsilon^{2a}$, after the complex calculations we shall gain expression for Airy function

$$\frac{d}{dt}\ln\psi^2 P^* = \frac{\Delta S}{K(\omega,x,y)K'_\omega(\omega,x,y)} - \frac{\partial^*}{\partial t}\ln K(\omega,x,y)K'_\omega(\omega,x,y) \qquad (4.9)$$

$$\nabla^* = \nabla + \frac{\partial}{\partial\omega}\nabla\omega, \qquad \frac{d}{dt} = \frac{\partial}{\partial t} + \frac{\mathbf{k}\nabla}{K(\omega,x,y)K'_\omega(\omega,x,y)}$$

$$\frac{\partial^*}{\partial t} = \frac{\partial}{\partial t} + \frac{\partial}{\partial\omega}\frac{\partial\omega}{\partial t}$$

$$P^* = \sqrt{\sigma(\omega,x,y)}\ K^{-3}(\omega,x,y)$$

where $d/dt$ is the derivative along the characteristics (4.8). For Fresnel wave $P^* = K^{-1}(\omega,x,y)$. Further we shall have

$$\frac{\Delta S}{K(\omega,x,y)K'_\omega(\omega,x,y)} - \frac{\partial^*}{\partial t}\ln K(\omega,x,y)K'_\omega(\omega,x,y) =$$

$$= -\frac{\partial^*}{\partial t}\ln K(\omega,x,y)K'_\omega(\omega,x,y) + \frac{\nabla S}{K(\omega,x,y)K'_\omega(\omega,x,y)}\nabla^*\ln K(\omega,x,y)K'_\omega(\omega,x,y) -$$

$$-\frac{\nabla S}{K(\omega,x,y)K'_\omega(\omega,x,y)}\nabla^*\ln K(\omega,x,y)K'_\omega(\omega,x,y) + \frac{\Delta S}{K(\omega,x,y)K'_\omega(\omega,x,y)} =$$

$$= -\frac{d}{dt}\ln K(\omega,x,y)K'_\omega(\omega,x,y) + \frac{\nabla S}{K(\omega,x,y)K'_\omega(\omega,x,y)}\nabla^*\ln K(\omega,x,y)K'_\omega(\omega,x,y) +$$

$$+\frac{\Delta S}{K(\omega,x,y)K'_\omega(\omega,x,y)} =$$

$$= - \frac{d}{dt} \ln K(\omega,x,y) K_\omega^{'}(\omega,x,y) - \nabla^*(-\frac{\nabla S}{K(\omega,x,y) K_\omega^{'}(\omega,x,y)}) =$$

$$= - \frac{d}{dt} \ln K(\omega,x,y) K_\omega^{'}(\omega,x,y) - \nabla^* \mathbf{c}$$

where $\mathbf{c}$ is the vector of the group velocity of the internal waves. Further we shall take advantage of the theorem of Liouville claiming, that along the characteristics (4.8) the ratio $dD(t,t_0,l)/dt = \nabla^* \mathbf{c}$ will be true. Then (4.9) may be written as a kind power conservation law along the ray

$$\frac{d}{dt} \ln(D\psi^2 P) = 0 \qquad (4.10)$$

where $P = K(\omega,x,y) K_\omega^{'}(\omega,x,y) P^*$

From here, considering, that all functions depend on ray coordinates, we shall gain

$$D(t,t_0,l)\psi^2(t,t_0,l) P(t,t_0,l) = E(l,t_0) \qquad (4.11)$$

where $E(l,t_0)$ is some function, which one cannot be determined, if we are solving the problem by the ray optics method. We shall note, that the power conservation law (4.10) similarly to the case of locally-harmonious waves, one may write the following way

$$\frac{\partial^* I}{\partial t} + \nabla^*(I\mathbf{c}) = 0$$

where $I = \psi^2 P$ is some "adiabatic variant "of a corresponding wave.

For the final solution of the problem it is necessary to determine the function $E(l,t_0)$ in the equation (4.11), which one can be find, using ноу solution of the problem about motion of a point source of mass in the stratified, horizontally-homogeneous medium, as on the typical distances from the source, where the global asymptotics is true in the horizontally-homogeneous medium (the order of several $h$) it is possible w to suppose, that the parameters of the medium characterizing the horizontal inhomogeneity varies a little, that means it is possible to consider the medium as the locally-homogeneous in horizontal direction. Let the source moving at the velocity $V$, in the moment of time $t = t_0$ is in the point $(x_0, y_0) = (Vt_0, 0)$. In each moment of time $t_0$ the source emits the waves of all frequencies over the range $0 < \omega < \max N(z)$, so as the ray coordinates it is convenient to take $\omega$ and the moment, when the ray is leaving the source $t_0$.

Then for Airy wave we shall gain $E(\omega, t_0) = \dfrac{\omega^4}{2V K^3(\omega, x_0, y_0) \nu(\omega, x_0, y_0)} \left( \dfrac{\partial f(x_0, y_0, z_0, \omega)}{\partial z_0} \right)^2$

$$\nu(\omega, x_0, y_0) = \sqrt{K^2(\omega, x_0, y_0) - \omega^2 V^{-2}}$$

For Fresnel wave the function $E(\omega, t_0)$ looks like

$$E(\omega, t_0) = \dfrac{\omega^2}{2V K(\omega, x_0, y_0) \nu(\omega, x_0, y_0)} \left( \dfrac{\partial f(x_0, y_0, z_0, \omega)}{\partial z_0} \right)^2$$

We shall write out the first term of the asymptotics of the vertical velocity component of the global Airy wave formed at motion of a point source of mass in the stratified horizontally-heterogeneous medium

$$w = w_0(t, t_0, \omega) f(x, y, z, \omega) \, \text{Ai}'\left( \left[ \dfrac{3}{2} S^*(t, t_0, \omega) \right]^{2/3} \right) \tag{4.12}$$

$$w_0(t, t_0, \omega) = \dfrac{\omega^2 K(\omega, x, y) \sigma^{1/4}(\omega, x, y)}{[2 V K_\omega'(\omega, x, y) D(t, t_0, \omega) K^3(\omega, x_0, y_0) \nu(\omega, x_0, y_0)]^{1/2}} \dfrac{\partial f(x_0, y_0, z_0, \omega)}{\partial z_0}$$

where $x = x(t, t_0, \omega)$, $y = y(t, t_0, \omega)$. The first term of the asymptotics of elevation of the global Fresnel wave looks like

$$\eta = \eta_0(t, t_0, \omega) f(x, y, z, \omega) \, \Phi\left( \sqrt{2 S^*(t, t_0, \omega)} \right) \tag{4.13}$$

$$\eta_0(t, t_0, \omega) = \dfrac{\omega}{[2 V K_\omega'(\omega, x, y) K(\omega, x_0, y_0) \nu(\omega, x_0, y_0)]^{1/2}} \dfrac{\partial f(x_0, y_0, z_0, \omega)}{\partial z_0}$$

The gained solutions (4.12) - (4.13) in the horizontally-homogeneous medium coincide with the global asymptotic forms built in [28]. At the great values of $S^*$ (far from the wave front), using Airy asymptotic forms and Fresnel integrals at the great values of the argument, we shall gain normal WKB expansion of the locally-harmonious waves for the horizontally-heterogeneous medium, and at the small values of $S^*$ (near to the wave front) we shall gain the solution in approximation of the week dispersion. Thus, the built solutions in the most general view describe the field of the internal waves formed at motion of the source of mass in the stratified horizontally heterogeneous medium.

Now it is interesting to compare now the analytical results gained above to the analysis of measurings of variability of the internal waves in the real medium with the varying in horizontal direction characteristics, particularly in the northwest area of the Pacific ocean, according to the data gained by the buoy stations during "Megapolygon" domestic experiment in the northern area of the Pacific ocean. Measurings of the flows and temperatures by Megapolygon" buoy stations has allowed to study the variability of the tidal internal waves on the area of 460 x 520 km. Measurings were conducted on all 170 buoys of the experiment on the horizon of 120 m, and on the horizon of 1200 m there have been used the additional measuring sets of temperature on buoys in the northwest area (14 buoys) and in the southeast area (16 buoys) of the polygon (Fig. 4).

Calculation of the estimates of the time-space spectrums for evaluation of parameters (arguments) of the internal waves has been conducted in compliamce with the method offered by Barber. Having the measurings of the temperature conducted in several points of the ocean on one horizon we can calculate the mutual spectrums of oscillations P and Q (so-called cospectrum and the quadrature spectrum). Knowing all the mutual spectrums and setting the frequency of the semidiurnal rising tide $f_0$, it is possible to make transformation to determine{the distribution of the mutual spectral power at this frequency on the wave numbers $k_x$ and $k_y$ (where $k^2 = k_x^2 + k_y^2$):

$$E(k_x, k_y, f_0) = \int_{-\infty}^{\infty} \int_{-\infty}^{\infty} [P(X, Y, f_0) - iQ(X, Y, f_0)] \exp[-2\pi i(k_x X + k_y Y)] dX dY$$,

where X and Y are the space intervals along the axes x и y.

As instead of the continuous spectrum of the space intervals we have the final set of the certain lengths corresponding to the space intervals between the buoy stations, then the integration is reduced to summation, and the calculation of the space spectrums for the frequency $f_0$ was conducted by the following formula

$$E(k_x, k_y, f_0) = 2\sum_{i=1}^{n-1} \sum_{i+1}^{n} [P_{ij}(f_0) \cos 2\pi(k_x x_{ij} + k_y y_{ij})] - [Q_{ij}(f_0) \sin 2\pi(k_x x_{ij} + k_y y_{ij})]$$.

Here k is the wave number, at that the wavelength $L = 1/k$; i, j is the numbers of the buoy stations; n is the quantity of the buoys; $x_{ij} = x_i - x_j$ and $y_{ij} = y_i - y_j$ are the projections of the space intervals between the buoys i and j on the axes x and y.

As the arrangement of the buoy stations in the northwest and southeast clusters of the polygon has been realized on the almost regular dislocation network, both clusters unambiguously allow the waves in the range from 77 up to 400 km, that well suits for the tidal internal waves with the length from 100 up to 150 km.

The length of the tidal internal wave was calculated by integration of the equation for the vertical velocity w in the internal wave at the real distribution of Brunt-Väisälä frequency in depth and the null boundary conditions for the ocean surface and bottom:

$$\frac{d^2 w}{dz^2} + \frac{N^2(z)}{g} + \frac{N^2(z) - \omega^2}{\omega^2 - f^2} k^2 w = 0$$.

The wavelength so calculated for the first mode in the area of "Megapolygon" was equal to about 130 km and near to the Imperial ridge the wavelength becomes bigger and is equal to 167 km, and 2000 km to the east it is equal to 156 km. In the Pacific north-east area the data of the buoy stations arranged by the American oceanologists were analyzed. The diagram of all buoys location is shown on Fig.5. The buoy stations of the north-west cluster had the time period of the synchronous operation from August, 10th till October, 7th, 1987, and on the south-east cluster such operation period was from September, 22-nd till October, 19th. From these realizations we selected the intervals with duration of 12-13 days for calculations of the estimates of the tidal internal waves parameters.

Thus for the south-east cluster two independent intervals have been selected, and for north-west cluster – four independent intervals. All the special spectral estimates gained in compliance with the independent data have the common maximum on the wavelength characteristic for the first mode of the internal tide and lay within the limits of from 110 up to 150 km at a mean of

130 km. The direction of the waves propagation is also rather stable and lays within the limits of 240-300 degrees, that corresponds to the propagation of the waves in the western and north-western directions from the Imperial ridge. At propagation of the tidal internal wave along the offshore water of "Megapolygon" some diffraction has been observed, that is the change of the direction of the wave propagation from the north-western direction in the south-east part of the polygon to the western direction in its north-western part.

Now we shall consider variation of the amplitudes of the internal tidal waves at itheir propagation in the western and the eastern directions from Imperial mountains. The amplitudes of the internal waves were calculated according to the deflection of the measured values of the temperature at the buoy stations from a mean with the subsequent division of this value by the average vertical gradient of temperature. The measurements were conducted on the horizon of 1200 m, and on the American buoys - to the east from the ridge on the horizon of 670-680 m. Variations of the amplitudes of the internal tidal waves inflow depending on the spacing intervals are shown on the Fig. 6. The calculations indicate, that the amplitude of the waves is decreasing approximately by 10 % at the distance equal to one length of the tidal internal wave (130-150 km).

We shall explore the influence of the various factors, including the heterogeneity in the orizontal direction, on attenuation of the internal waves. Within the limits of the above presented theory we shall consider the evolution of the internal gravity wave from the frequency $\omega$ corresponding to the semidiurnal period $T = 12$ hours. At that the slow variation of stratification along the path of the wave propagation is assumed. On the basis of the real geometry of the experiment it is supposed, that the considered problem is two-dimensional, that is the stratification depends only on the two variables: the depth $z$ and the distance from the path of the wave propagation $x$.

Now we shall consider the case of he constant depth $H$ and the stratification $N$, linearly depending only on $x$: $N(x) = N_1 + (N_2 - N_1)x/L$, where $L$ is the distance between two observation points; $x = x_1 = 0$ is the initial point; $x = x_1 = L$ is the terminal point; $N_{1,2} = N(x_{1,2})$. We shall consider only the first mode $\eta_1(z,x)$ of the amplitude of the vertical displacement of particles, neglecting its index. For solution of the given problem we shall use one of the modifications of the ray optics method "vertical modes - horizontal beams". In this case the amplitude $\eta(z,x)$ is searched in the form of

$\eta(z,x) = A(x) f(z,x)$,

where $f(z,x)$ - the normalized eigenfunction of the standard boundary-value problem of the equation of internal waves with the normalization

$$\int_0^H (N^2(x) - \omega^2) f^2(z,x) dz = 1$$

which one looks like $f(z,x) = \sqrt{\dfrac{2}{H(N^2(x) - \omega^2)}} \sin(\pi z / H)$.

The amplitude $A(x)$, depending only on $x$, is determined from the following conservation law:

$$\dfrac{A^2(x_1)}{k^2(x_1)} da(x_1) = \dfrac{A^2(x_2)}{k^2(x_2)} da(x_2),$$

where $k(x)$ is the horizontal wave vector magnitude; $da(x)$ - the width of an elementary wave tube. As the problem is two-dimensional, then the width of the ray tube does not vary along the ray and the conservation law becomes simpler: $A(x)/k(x) = \text{const}$. Due to the of the smallness of the considered values of $\omega$, the velocity of the wave propagation is close to the maximal group velocity $c(x) = N(x)H/\pi$, therefore the wave number $k(x) = \omega\pi/N(x)H$ and the corresponding to it wavelength $\lambda(x) = 2N(x)H/\omega$. Then, assuming, that the depth of the observation points is similar, we shall have from the conservation law $(A_{1,2} = A(x_{1,2}))$: $A_1 N_1 = A_2 N_2$, or $A_2 = A_1 \lambda_1 / \lambda_2$. Then the values for the complete amplitude look like

$$W_{1,2} = A_{1,2} \sqrt{\frac{2}{H(N_{1,2}^2 - \omega^2)}}.$$

From here we have

$$W_2 = W_1 \frac{N_1}{N_2} \sqrt{\frac{(N_1^2 - \omega^2)}{(N_2^2 - \omega^2)}}$$

Or as $\omega \ll N$, then $W_2 = W_1 \lambda_1^2 / \lambda_2^2$, that is the amplitude of the internal wave is inversely proportional to a square of the wavelength.

The wave travel time $\tau$ along the horizontal ray is determined from the equation of characteristics $\frac{dx}{dt} = c(x)$, $c(x) = (N_1 + ax)H/\pi$, $a = (N_2 - N_1)/L$.. Integrating this equation, we determine the travel time of the wave

$$\tau = \frac{\pi}{aH} \ln\left(\frac{N_2}{N_1}\right) = \frac{TL}{(\lambda_2 - \lambda_1)} \ln\left(\frac{\lambda_2}{\lambda_1}\right). \tag{4.14}$$

The available field data present the following values of the basic parameters of the problem: $\lambda_1 = 167$ km, $\lambda_2 = 156$ km, $L = 2000$ km. The attenuation factor without consideration of the wavelength describing the degree of the amplitude decreasing on the wavelength and further designated through $\beta$, in view of the known from the results of observations ratio $W_2 / W_1 = 0{,}2 \equiv \beta^{t/T} = \beta^{L/\lambda}$ gives the values for $\beta : \beta = 0{,}2^{167/2000} = 0{,}874$. The attenuation in view of variation of the wavelength along the ray is determined by: $W_2 / W_1 = \beta^{\tau/T}$, where the value of $\tau$, is determined from the ratio (4.14), gives the value for $\beta = 0{,}878$.

Thus, the gained estimates allow to assert, that the described in this paragraph method of asymptotic introduction of the wave fields allows to consider the influence of the wave field density heterogeneities, which is one of the factors determining the free-space attenuation of the internal gravity waves field in the full scale measurements.

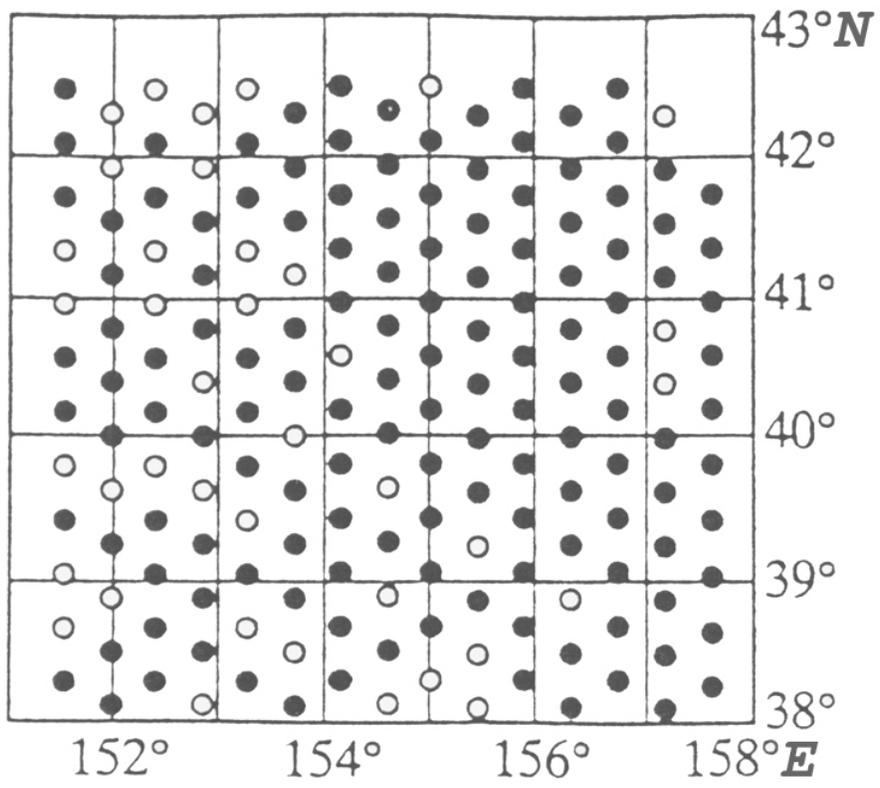

Fig.4

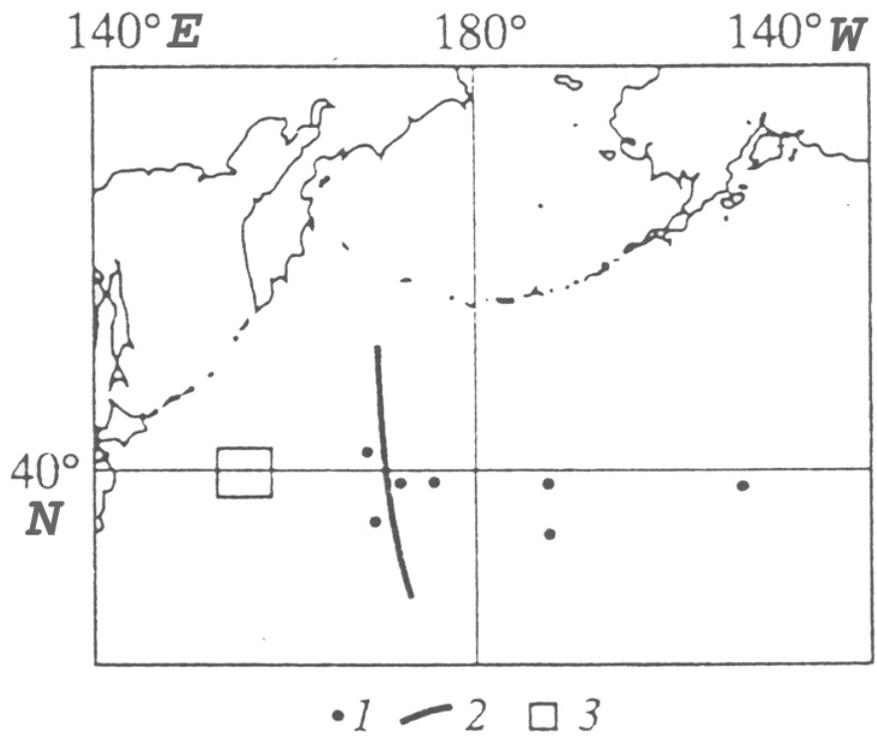

Fig.5

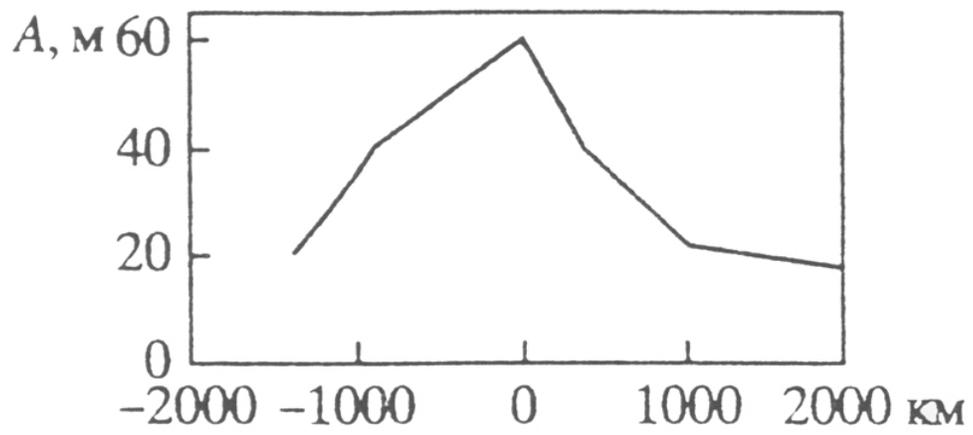

Fig.6

# 5. Local asymptotics of the far field of the internal gravity waves in the stratified heterogeneous in the horizontal direction medium

The present section using the method of the "traveling wave" being generalization of the ray optics method is solving the problem of propagation of Airy and Fresnel waves in the medium with heterogeneous in the horizontal direction field of density and shall consider only the field near to the wave front, that is it will use the weak dispersion approximation. As well as in the previous Sections it is considered, that the scale of the density variation in the horizontal direction is much bigger than the reference lengths of the internal waves, the slowness of variation in the vertical direction is not supposed.

As Airy wave in the present paragraph we shall mean the following asymptotic solution of the equation of the internal waves for elevation $\eta$ in the horizontally-homogeneous stratified layer $0 < z < -H$ with Brunt-Väisälä frequency $N^2$

$$\eta = Q(3\beta_n l)^{-1/3} \text{Ai}\left[\frac{t - \sigma_n l}{(3\beta_n l)^{1/3}}\right] f_n(z), \qquad (5.1)$$

where $\text{Ai}(x)$ is Airy function Эйри, $f_n(z)$ and $\sigma_n$ are eigenfunctions and eigenvalues of the problem

$$f_n'' + \sigma_n^2 N^2(z) f_n = 0, \quad f_n = 0 \quad (z = 0, -H). \qquad (5.2)$$

Coefficients $\sigma_n$, $\beta_n$ are the first two coefficients of expansion values of the dispersion curve $k_n(\omega) = \sigma_n \omega + \beta_n \omega^3 + \Lambda$,

$$\beta_n = \int_{-H}^{0} f_n^2(z) dz \left[\int_{-H}^{0} N^2(z) f_n^2(z) dz\right]^{-1}.$$

Here $l = x \cos\Theta + y \sin\Theta$, $\Theta$ is the angle of the wave expansion with the axis $x$.

As Fresnel wave we shall mean the following asymptotic solution of the equation of the internal waves for elevation $\eta$ in the horizontally-homogeneous stratified layer $0 < z < -h$ with Brunt-Väisälä frequency $N^2(z)$, lying on the homogeneous layer of the infinite depth with $N^2(z) = 0$

$$\eta = \text{Re}\left(P(\gamma_n l)^{-1/2} \Phi^*\left[\frac{t - \lambda_n l}{(\gamma_n l)^{1/2}}\right] \psi_n(z)\right), \qquad (5.3)$$

$$\Phi^*(\sigma) = \int_0^\infty \exp\left(-it\sigma - i\frac{t^2}{2}\right) dt \equiv F_0(\sigma),$$

where $\psi_n(z)$ and $\lambda_n$ are eigenfunctions and eigenvalues of the problem

$$\psi_n'' + \lambda_n^2 N^2(z) \psi_n = 0, \quad \psi_n = 0 \quad (z = 0), \quad \psi_n' = 0 \quad (z = -h). \qquad (5.4)$$

Coefficients $\lambda_n$, $\gamma_n$ are the first two coefficients of expansion of the dispersion curve $k_n(\omega) = \lambda_n \omega + \gamma_n \omega^2 + \Lambda$,

$$\gamma_n = \psi_n^2(-h) \left[\int_{-h}^{0} N^2(z) \psi_n^2(z) dz\right]^{-1}.$$

Amplitude factors $Q, P$ are determined by the particular kind of the being solved problem..
Further we shall consider one separately taken mode, neglecting the index $n$.

$$\frac{\partial^2}{\partial t^2}\left(\Delta + \frac{\partial^2}{\partial z^2}\right)\eta = g \,\Delta(\varphi \,\text{grad}\ln\rho), \qquad (5.5)$$

$$\Delta = \frac{\partial^2}{\partial x^2} + \frac{\partial^2}{\partial y^2},$$

where the velocity vector $\mathbf{u} = \{u, v, w\}$ is expressed through the potential $\varphi = \{\varphi_1, \varphi_2, \eta\}$ in the following way $\mathbf{u} = \partial \varphi / \partial t$.

The solution for the elevation is determined similarly as in Sections 3.2-3.4 and it is searched as follows

$$\eta = \operatorname{Re} \eta^*, \tag{5.6}$$

$$\eta^* = A(\varepsilon x, \varepsilon y, z) F_0(r) + i\sqrt{\varepsilon}[B(\varepsilon x, \varepsilon y, z) F_1(r) + C(\varepsilon x, \varepsilon y, z) F_{-1}(r)] + O(\varepsilon),$$

where the argument $r = \alpha(\varepsilon x, \varepsilon y)[\varepsilon t - S(\varepsilon x, \varepsilon y)]\varepsilon^{-1/2}$ is considered of the order of unity, the phase functions as well as earlier are coupled by the ratios: $F_m(r) = F_{m-1}'(r)$, the argument $\varepsilon$ characterizes " slow variables ", function $S$ determines a wave front position at $r = 0$, function $\alpha$ describes the evolution of the width of the pulse of Fresnel wave. On the surface $z = 0$ the boundary value condition of the " solid cover" $\eta = 0$ is set, on border of the stratified layer $z = -h$ - the boundary condition ensuring the exponential fading of the solution with the increasing depth $|\partial \eta / \partial z| = |\nabla \eta|$, where $\nabla = \{\partial/\partial x, \partial/\partial y\}$.

Further it is necessary to express the first two components $\varphi_1, \varphi_2$ of potential $\varphi$ through elevation $\eta$. Using link between elevation $\eta$ and horizontal vector of velocity $\mathbf{u} = \{u, v\}$ in the form of

$$\Delta \mathbf{u} = \frac{\partial^2}{\partial t \partial z}(\nabla \eta),$$

In the first approximation with respect to $\varepsilon$ it is possible to gain

$$\varphi_h = -A_z'(\varepsilon x, \varepsilon y, z) \frac{\nabla S F_1(r)}{\alpha \sqrt{\varepsilon}} + O(1) \tag{5.7}$$

$$\varphi_h = \{\varphi_1, \varphi_2\}.$$

later on we shall be interested in the first term of the expansion (5.6). Substituting (5.6), (5.7) in the equation (5.5) and the boundary conditions, equating the terms at the equal degrees $\varepsilon$ and similar functions $F_m(r)$, we shall have

$$LA = 0, \tag{5.8}$$

$$i\alpha^2 LC = N^2(-2A\nabla\alpha\nabla S + \alpha A \Delta S + 2\alpha\nabla\alpha\nabla S), \tag{5.9}$$

$$i\alpha^2 F_3 LB = 2N^2 A(rF_2 + 2F_1)\nabla\alpha\nabla S,$$

$$A = B = C = 0 \quad (z = 0), \tag{5.10}$$

$$A_z' = C_z' = 0, \quad B_z' = A|\nabla S|\alpha \quad (z = -h),$$

Where $L = \partial^2 / \partial z^2 + N^2 |\nabla S|^2$, $N^2 = N^2(x, y, z)$.

As $F_3 = irF_2 + i2F_1$, then from (5.10) we shall gain

$$LB = -2N^2 A \alpha^{-2} \nabla \alpha \nabla S. \tag{5.11}$$

At the solution of this problem it is supposed, that the eigenfunctions and eigenvalues of the problem (5.4) are known. Then for determination of function $S$ we shall write down the equation of the eikonal with its known right part

$$|\nabla S| = \lambda(x, y) \equiv c^{-1}(x, y) \tag{5.12}$$

For determination of functions $A$, $\alpha$ it is necessary to multiply the equations (5.9), (5.11) by $A$ and to integrate them from $-h$ up to 0. Using the boundary conditions and the horizontal properties of the problem (5.8), we shall gain

$$\nabla S \nabla \ln(\Phi c \alpha^{-2}\rho^{-\chi}) + \Delta S = 0, \tag{5.13}$$

$$\gamma \alpha^3 + 2c\nabla\alpha\nabla S = 0, \tag{5.14}$$

$$\Phi = \int_{-h}^{0} N^2 A^2 dz, \qquad \chi = \frac{1}{2}g\gamma.$$

For solving the equation of the eikonal (5.12) we shall make out the system of characteristic equation

$$\frac{dx}{dt} = c^2(x,y)p, \qquad \frac{dp}{dt} = -\frac{c'_x(x,y)}{c(x,y)},$$

$$\frac{dy}{dt} = c^2(x,y)q, \qquad \frac{dq}{dt} = -\frac{c'_y(x,y)}{c(x,y)}. \tag{5.15}$$

With the characteristics of (5.15) the equation (5.13) is reduced to the following conservation law:

$$\frac{d}{d\tau}(D\Phi\rho^{-\chi}\alpha^{-2}c^{-1}) = 0, \tag{5.16}$$

where $d/d\tau$ is a derivative along the characteristics (5.15), $D$ is Jacobian of conversion from the usual coordinates to the ray coordinates $\tau, \tau_0$. The Jacobian $D$ can be expressed through the geometrical ray divergence of the ray tube $R$ : $D = cR$.

The equation (5.14) is solved using characteristics of (5.15)

$$\alpha = \left[\int_{\tau_0(x,y)}^{\tau(x,y)} \gamma c d\tau\right]^{-1/2}. \tag{5.17}$$

Now we shall consider Airy wave propagation. In this case the solution for elevation $\eta$ is searched in the following form

$$\eta = A(\varepsilon x, \varepsilon y, z) F_0(r) + \varepsilon^{2/3} C(\varepsilon x, \varepsilon y, z) F_{-1}(r) + O(\varepsilon^{4/3}) \tag{5.18}$$

Hereinafter all the indications are saved, but $F_0(r) = Ai(r)$. As the boundary conditions in the case we use approximation of "the solid cover" $\eta = 0$ on the surface and on the bottom. Substituting the solution in the form of (5.18) into the equation (5.5) and iterating the procedure set forth above we shall gain the eikonal equation for determination of function S

$$|\nabla S| = \sigma(x,y), \tag{5.19}$$

where function $\sigma(x,y)$ is determined from the problem (5.2).

On the characteristics of the eikonal equation (5.19) the conservation law is executed

$$\frac{d}{d\tau}(D\Phi\alpha^{-2}\sigma) = 0 \tag{5.20}$$

The function $\alpha$, describing the evolution of Airy wave pulse width is searched by integration along the characteristics

$$\alpha = \left[3 \int_{\tau_0(x,y)}^{\tau(x,y)} \beta d\tau\right]^{-1/3}. \tag{5.21}$$

For the numeric calculation we used the following model close to the real parameters of the wavetrains of the internal gravity waves at ocean. The boundary of the heterogeneity passes along the axis on a fulcrum $y$, on which at an angle of 45° the section of Fresnel plane wave of 750 m length (see Fig..7), the law of density variation is set in the following form $\rho = \exp[N_0^2 z(1+\varepsilon x)/g]$, $N^2 = N_0^2(1+\varepsilon x)$, where $\varepsilon = 0{,}0016$ $N_0 = 0{,}01\,c^{-1}$, the stratified layer thickness $h$ of $=100$ m.

Then for everyone separate $n$-th mode we shall have

$$k_n(\omega) = \lambda_n \omega + \gamma_n \omega^2 + \Lambda$$

$$c_n(x) = 1/\lambda_n(x) = \frac{N_0 h\sqrt{1+\varepsilon x}}{\pi(n-1/2)}$$

$$\frac{\partial c_n(x)}{\partial x} = \frac{\varepsilon N_0 h}{2\pi\sqrt{1+\varepsilon x}(n-1/2)}$$

Then the propagation velocity of the plane wave in the homogeneous medium is $c_n = 2N_0 h/\pi(n-1/2)$.

Let's introduce further the amplitude factor in the following form $A_n(z,x) = \psi_n(z,x)\varphi_n(x)$
where the eigenfunctions $\psi_n(z,x)$ such as

$$\int_0^h \psi_n^2(z,x) N^2(z,x) dz = 1$$

Have the form

$$\psi_n(z,x) = \sin(\frac{\pi z}{h}(n-1/2))\sqrt{2/h}\left(N_0\sqrt{1+\varepsilon x}\right)^{-1}$$

The second factor of expansion of the dispersion curve describing the spatial evolution of Fresnel wave pulse width looks like

$$\gamma_n = \left(N_0^2(1+\varepsilon x)h(n-1/2)\right)^{-1}$$

The conservation law along the characteristics looks like

$$\frac{R\varphi_n^2}{\rho^\chi \alpha_n^2} = const$$

$$\rho^\chi = \exp(\frac{z}{2h(n-1/2)})$$

$$\alpha_n = \left[\int_0^\tau \gamma_n c_n d\tau\right]^{-1/2}$$

Because in this case the value of $\rho^\chi$ does not depend on the horizontal coordinates the conservation law along the characteristics has a more simple form

$$\frac{R\varphi_n^2}{\alpha_n^2} = const \qquad (5.22)$$

The surface of the planar front of Fresnel incident wave for the given geometry of the problem is described by the following function

$$S(t,x,y) = t - \frac{x+y}{\sqrt{2}\,c_n}$$

Proceeding from the known formula of the normal velocity $n$ of propagations of the $S(t,x,y)$ surface in the form of

$$n = \frac{\partial S}{\partial t}\left(\left(\frac{\partial S}{\partial x}\right)^2 + \left(\frac{\partial S}{\partial y}\right)^2\right)^{-1}$$

we gain, that the surface of the wave front is propagating at the velocity of $c_n$.

Then let's determine the value of the constant in (3.5.22). The geometrical divergence of the rays, evidently is the ratio of the width of the ray tube $J_2$ in the end of the ray to the width of the ray tube $J_1$ in the point of the ray outlet

$$R = \frac{J_2}{J_1}$$

And as in the considered case the rays are propagating in the plane *(x,y)*, then the width of the ray tube is the spacing interval between two near by rays. In the point of the ray outlet, when the effect of the horizontal heterogeneity is negligibly small, the rays are almost in parallel to each other and, accordingly, the geometrical rays divergence $R_1$ in the point of the ray outlet is equal to unity,

$$\eta = \text{Re}\left((\gamma_n l)^{-1/2}\, \Phi^*\left[\frac{t - \lambda_n l}{(\gamma_n l)^{1/2}}\right]\psi_n\right) = \text{Re}\left(\varphi_n \psi_n \Phi^*(\alpha_n(t-S))\right)$$

we gain, that

$$\varphi_n = \alpha_n$$

and the value of the constant in (5.22) is also equal to unity. Thus, the incidence of the planar section of Fresnel wave at an angle of 45° to the boundary of the heterogeneous medium passing along the axis of ordinates one may physically interpret as the motion of some source along this axis vertically upward and emitting the rays, which are propagating in the horizontally heterogeneous medium.

The Fig. 7 presents the evolution of the given section of the planar wave front of Fresnel wave; the Fig. 8 indicates the results of calculation of the wave in A and B points. One can see, that the horizontal heterogeneities of the field of density results in the significant variation of the wave amplitude and in the time shift being determined by the curvature of the rays

Thus, the solution of the problem about Airy and Fresnel waves propagation in the stratified heterogeneous in the horizontal direction medium is reduced to the solution on the characteristics of the eikonal equations (5.12), (5.19). Thus it allows us to determine the wave front position. The law of variation the wave amplitude is gained from the equations (5.16), (5.20). Variation of the pulse width of Airy and Fresnel waves can be calculated from the following equations (5.17), (5.21). According to the results of the conducted numerical calculations aimed at determination of the real scale of the horizontal variability of the field density, at presence of such horizontal heterogeneities there take place the significant curvature of the wave fronts, as well as variations of the width and amplitude of the pulses of the inharmonic wave trains. In the considered case it is true for Airy and Fresnel waves.

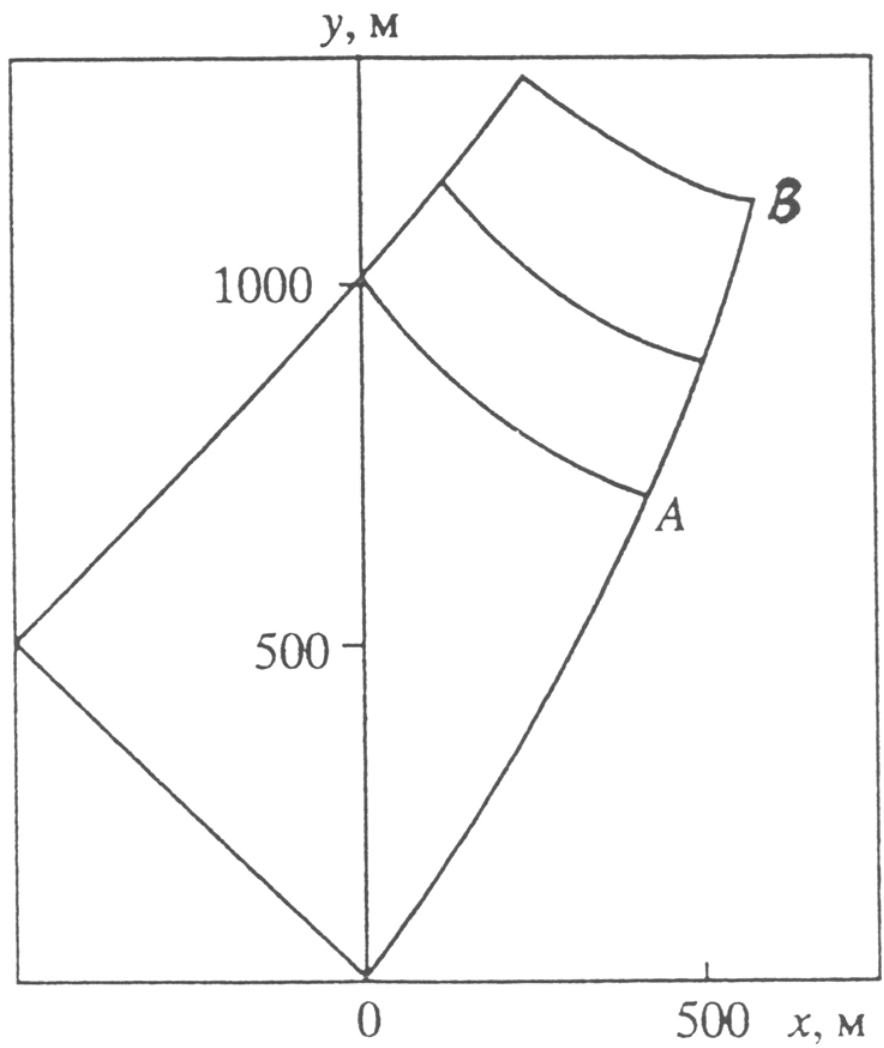

Fig.7

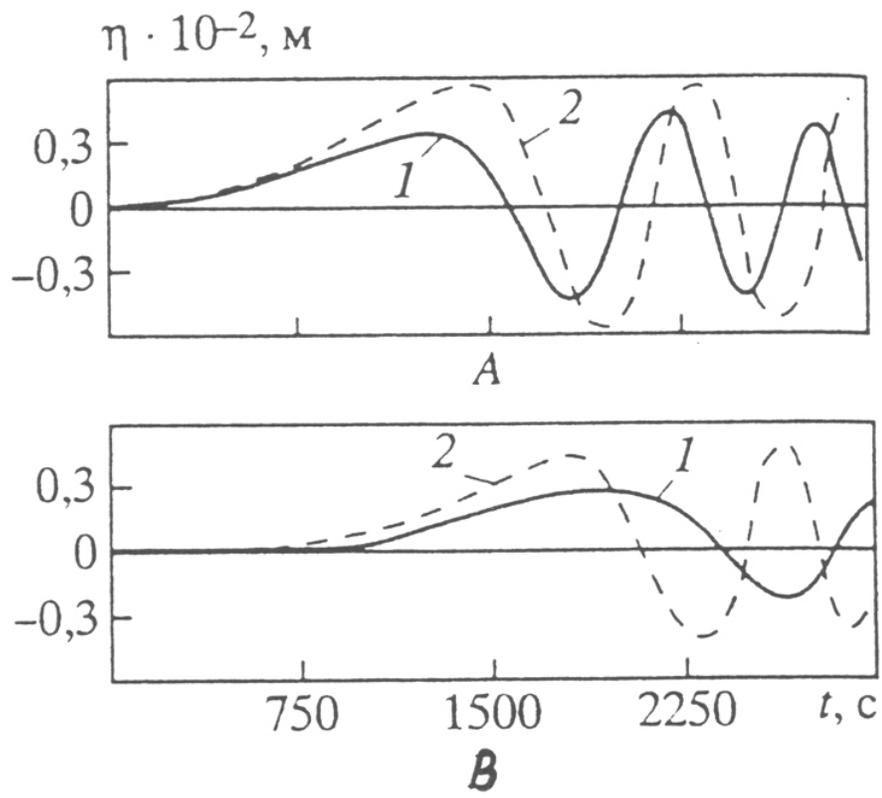

Fig.8

# 6. Local asymptotics of the far field of the internal gravity waves in the stratified non-stationary medium

In the present section we using the method of the "traveling wave" have solved the problem of propagation of Airy and Fresnel internal waves in the non-stationary stratified mediums. In the real ocean conditions Brunt-Väisälä frequency ( $N^2 = -g\partial \ln\rho/\partial z$, where g - gravity acceleration, $\rho$ - the undisturbed density defining the basic characteristics of the internal gravity waves) depends not only on the spatial variables $x, y, z$, but also on the time $t$. The most typical kinds of variability of $N^2$ with respect to the time are raising and lowering of the thermocline, variation of its{ width etc. There are some time scales of variability of the hydrophysical fields at oceans and in seas : the small-scale variability with the time periods up to ten minutes, the mesoscale variability with the time periods up to 24 hours, and also the synoptic variability and the global variability with the time periods from months up to several years. Later we shall consider propagation of the internal gravity waves in the non-stationary mediums, which parameters have the variation periods of 24 hours and more, that allows to use the approximation of the ray optics, as the internal waves period makes dozens of minutes and less.

The system of the linearized hydrodynamical equations, when the undisturbed density $\rho$ depends on variables z фтв t, is reduced to one equation, for example for the vertical velocity

$$\left(\frac{\partial}{\partial t} + \frac{\partial \ln\rho}{\partial t}\right)\left[\frac{\partial}{\partial t}\left(\Delta + \frac{\partial^2}{\partial z^2}\right) + \frac{\partial \ln\rho}{\partial z}\frac{\partial^2}{\partial t\partial z}\right]w = g\frac{\partial \ln\rho}{\partial z}\Delta w$$

where $\Delta = \frac{\partial^2}{\partial x^2} + \frac{\partial^2}{\partial y^2}$.

Neglecting the term with $\partial \ln\rho/\partial z$, we shall gain the equation in Boissinesq approximation

$$\left(\frac{\partial}{\partial t} + \frac{\partial \ln\rho}{\partial t}\right)\frac{\partial}{\partial t}\left(\Delta + \frac{\partial^2}{\partial z^2}\right)w + N^2(z,t)\Delta w = 0$$

It seems natural to neglect as well the term with $\partial \ln\rho/\partial t$, that would correspond to the successive application of Boissinesq hypothesis . It means, that the density characterizing the inertial mass of the liquid, can be considered as a stationary value. Then

$$\frac{\partial^2}{\partial t^2}\left(\Delta + \frac{\partial^2}{\partial z^2}\right)w + N^2(z,t)\Delta w = 0. \tag{6.1}$$

The gained equation differs from the normal equation of the internal waves in the stationary medium only by its parametric time $t$ introduced in Brunt-Väisälä frequency .

The solution (6.1) is searched in the form of the sum of the modes, each of which is propagating irrespective of others (adiabatic approximation). We shall consider one separately taken mode neglecting its index. Further we shall be interested in the area near to the wave front, that is we shall consider only time $t$, which is close to the time of coming of the wave front, which will be expressed later on by $\tau$, that is we shall use slightly dispersive approximation.

We shall consider Airy wave propagating in the layer of the stratified liquid $-H < z < 0$ with Brunt-Väisälä frequency $N^2(z,t)$. We shall search for the solution (3.6.1) with the boundary conditions $w = 0$, $z = 0, -H$ in the form of

$$w = \left[ A(\varepsilon x, \varepsilon y, \tau, z) + \frac{\partial A(\varepsilon x, \varepsilon y, \tau, z)}{\partial \tau}(\varepsilon t - \tau) + \Lambda \right] F_0(\varphi) +$$
$$+ \varepsilon^p \left[ B(\varepsilon x, \varepsilon y, \tau, z) + \frac{\partial B(\varepsilon x, \varepsilon y, \tau, z)}{\partial \tau}(\varepsilon t - \tau) + \Lambda \right] F_1(\varphi) - O(\varepsilon^{2p}),$$
(6.2)

$$F_{m+1}'(\varphi) = F_m(\varphi),$$

where $p = 2/3$; $\tau = \tau(\varepsilon x, \varepsilon y)$; $F_0(\varphi) = Ai'(\varphi)$ is derivative of Airy function, which parameter $\varphi = \alpha(\varepsilon x, \varepsilon y)(\varepsilon t - \tau(\varepsilon x, \varepsilon y))\varepsilon^{-p}$ is of the order of unity. Function $\tau$ describes the position of the wave front, the function $\alpha$ describes the evolution of the width of Airy wave, the small parameter $\varepsilon$ characterizes the "slow variables".

As we are interested only in the "slow times" $\varepsilon t$, close to the time of incoming of the wave front $\tau$, then all the functions standing before the functions $F_m$, are presented in the form of Taylor series by powers $\varepsilon t - \tau \approx \varepsilon^p$.

Now we shall represent $N^2(z, \varepsilon t)$ in the following form

$$N^2(z, \varepsilon t) = N^2(z, \tau) + \frac{\partial N^2(z, \tau)}{\partial \tau}(\varepsilon t - \tau) + O(\varepsilon^{2p}).$$
(6.3)

Substituting expansions (6.2) and (6.3) in (6.1) and equating the terms with the equalexponents $\varepsilon$, we shall gain at $\varepsilon^p$

$$\frac{\partial^2 A}{\partial z^2} + |\nabla \tau|^2 N^2(z, \tau) A = 0,$$
(6.4)

$$A = 0, \quad z = 0, -H,$$
(6.5)

where $\nabla = (\partial/\partial x, \partial/\partial y)$. It is convenient to represent the eigenfunction of the problem (6.4) - (6.5) in the form of $A(x, y, \tau, z) = \Psi(x, y) f(z, \tau)$, where $f(z, \tau)$ meets the condition of normalizations

$$\int_{-H}^{0} N^2(z, \tau) f^2(z, \tau) dz = 1.$$

The eigenfunctions $f(z, \tau)$ and the eigenvalues $\lambda(\tau) \equiv |\nabla \tau|$ of the problem (6.4) - (6.5) are represented as already known, then for determination of $\tau$ we have the eikonal equation

$$\left(\frac{\partial \tau}{\partial x}\right)^2 + \left(\frac{\partial \tau}{\partial y}\right)^2 \equiv r^2 + q^2 = \lambda^2(\tau)$$
(6.6)

The corresponding characteristic system for the eikonal equation (6.6) looks like

$$\frac{dx}{d\tau} = rc^2(\tau), \quad \frac{dr}{d\tau} = -\frac{c'(\tau)}{c(\tau)} r,$$
$$\frac{dy}{d\tau} = qc^2(\tau), \quad \frac{dq}{d\tau} = -\frac{c'(\tau)}{c(\tau)} q.$$
(6.7)

Here $c(\tau) \equiv \lambda^{-1}(\tau)$ is the maximum velocity of the long waves.

For determination of functions $\alpha$ and $\Psi$ we shall first substitute expansions (6.2) and (6.3) in the equation (6.1) and then to equate the terms of the order $\varepsilon^{2p}$. We shall gain

$$\lambda^4(\tau) a(\tau) + \nabla \alpha \nabla \tau + \alpha \lambda^{-1}(\tau) \nabla \lambda(\tau) \nabla \tau = 0,$$
(6.8)

$$\nabla\left(\frac{\Psi^2 c^4}{\alpha^4}\right)\nabla\tau + \Delta\tau = 0,$$

$$a(\tau) = \int_{-H}^{0} f^2(z,\tau)\,dz.$$

For determination of $\Psi$ we shall have the following conservation law along the (6.7) characteristics:

$$\frac{\Psi^2 c^3 R}{\alpha^4} = \text{const}, \tag{6.9}$$

where the geometrical divergence of rays R is coupled with Jacobian D describing transition from the space variables x, y to the ray coordinates $\tau$ and $\tau_0$ by means of the expression $D = R c$.

On the characteristics of (6.7) the equation (6.8) is reduced to Bernoulli equation

$$\frac{d\alpha}{d\tau} + \frac{1}{\lambda(\tau)}\frac{d\lambda(\tau)}{d\tau}\alpha = -\alpha^4 a(\tau)$$

solution of which looks like

$$\alpha(x,y) = c(\tau)\left[\frac{3}{2}\int_{\tau_0(x,y)}^{\tau(x,y)} a(t) c^3(t)\,dt\right]^{-1/3}.$$

Now we shall consider Fresnel wave propagating in the layer of the stratified liquid of h depth with Brent-Вяйсяля frequency $N^2(z,t)$ and lying on the great homogeneous layer with $N^2(z,t) = 0$. We shall search for the solution, for example, of elevation $\eta$ ($w = \partial\eta/\partial t$). In the capacity of the boundary conditions we shall take $\eta = 0$ on the surface $z = 0$ and $|\partial\eta/\partial z| = |\nabla\eta|$ - on the boundary of the stratified layer $z = -h$. The last condition ensures an exponential attenuation of the solution in compliance with the varying depth. The solution for $\eta$ we shall search in the form of

$$\eta = \text{Re}\,\eta^*, \tag{6.10}$$

$$\eta^* = \left[A(\varepsilon x, \varepsilon y, \tau, z) + \frac{\partial A(\varepsilon x, \varepsilon y, \tau, z)}{\partial \tau}(\varepsilon t - \tau) + \Lambda\right] F_0(\varphi) +$$

$$+ i\varepsilon^p \left[B(\varepsilon x, \varepsilon y, \tau, z) + \frac{\partial B(\varepsilon x, \varepsilon y, \tau, z)}{\partial \tau}(\varepsilon t - \tau) + \Lambda\right] F_1(\varphi) +$$

$$+ i\varepsilon^p \left[C(\varepsilon x, \varepsilon y, \tau, z) + \frac{\partial C(\varepsilon x, \varepsilon y, \tau, z)}{\partial \tau}(\varepsilon t - \tau) + \Lambda\right] F_{-1}(\varphi) + O(\varepsilon^{2p}),$$

$$F_0(\varphi) = \int_0^\infty \exp\left(-i t \varphi - i\frac{t^2}{2}\right) dt,$$

where $p = 1/2$; all the rest symbols hereinafter are saved, except for the specially stipulated symbols. Substituting expansions (6.3) and (6.10) in the equation (6.1), we shall gain the eikonal equation (6.6) for determination of $\tau$, in which $\lambda(\tau)$ is determined from the equation (6.4) with the boundary conditions $A = 0, z = 0$, $\partial A / \partial z = 0, z = -h$. On the characterictics of the eikonal equation we have the following conservation law $\dfrac{\Psi c R}{\alpha^2} = \text{const}$.

Function $\alpha$ will be the solution of the corresponding Bernoulli differential equation

$$\alpha(x,y) = c(\tau) \left[ \int_{\tau_0(x,y)}^{\tau(x,y)} b(t) c^3(t) dt \right]^{-1/2},$$

$b(\tau) = f^2(-H, \tau)$.

Further let's consider Airy wave formed at movement of a point source of a mass in the fixed borders finite layer of stratified liquid. Let us assume, that the source moves in a positive direction of the axis x with the velocity V on the depth $z_0$. In the capacity of the parameter of integration of the system (6.7) we shall take the eikonal $\tau$. The result of the solution of the system (6.7) is the one-parameter family of functions $x(\tau, \tau_0)$, $y(\tau, \tau_0)$, $r(\tau, \tau_0)$, $q(\tau, \tau_0)$; the first two functions determine a ray on the plane x, y, $\tau_0$ is the initial eikonal or, that is the same, the time, when the ray leaves the source. Let in the moment of time $\tau = \tau_0$ the source is in the point $(x(\tau_0), y(\tau_0)) = (V\tau_0, 0)$, then for determination of the functions $r(\tau_0)$, $q(\tau_0)$ we shall gain the following system of the equations

$$r^2(\tau_0) + q^2(\tau_0) = c^{-2}(\tau_0), \tag{6.11}$$

$$\dfrac{dx(\tau_0)}{d\tau_0} r(\tau_0) + \dfrac{dy(\tau_0)}{d\tau_0} q(\tau_0) = 1. \tag{6.12}$$

The equation (6.11) is the eikonal equation (6.6) in the moment of time $\tau = \tau_0$, the equation (6.12) is gained by differentiation of the initial eikonal $\tau_0(x,y)$ with respect to $\tau_0$. The functions $r(\tau_0)$, $q(\tau_0)$ we shall gain from (6.11) - (6.12), the ratio of which $q(\tau_0)/r(\tau_0)$ determines the tangent of the angle between the ray, which is leaving the source in the moment $\tau = \tau_0$ from the point $(V\tau_0, 0)$, and the axis x. Then the initial data for the system (6.7) will be the following:

$$\begin{aligned} x(\tau_0) &= V\tau_0, & r(\tau_0) &= 1/V, \\ y(\tau_0) &= 0, & q(\tau_0) &= \sqrt{c^{-2}(\tau_0) - V^{-2}}. \end{aligned} \tag{6.13}$$

Solving the system (6.7) with the initial data (6.13), we shall gain the rays equations

$$\begin{aligned} x(\tau, \tau_0) &= V\tau_0 + \dfrac{c(\tau_0)}{V} \int_{\tau_0}^{\tau} c(t) dt \\ y(\tau, \tau_0) &= c(\tau_0)\sqrt{c^{-2}(\tau_0) - V^{-2}} \int_{\tau_0}^{\tau} c(t) dt \end{aligned} \tag{6.14}$$

From (6.14) follows, that the rays are direct lines, which inclination depends on the time, when the ray leaves the source $\tau_0$. At the fixed $\tau$ we shall have the wave front, at the fixed $\tau_0$ - the ray. Inverting the equations (6.14), we shall gain: $\tau = \tau(x,y)$, $\tau_0 = \tau_0(x,y)$. For the rectilinear and uniform motion of the point source of a mass it is possible to determine the value of the constant in the right member of the equation (3.6.9), in the further represented as $P(\tau_0)$. The value of $P(\tau_0)$ is searched from the problem with the constant in time Brunt-Väisälä stationary frequency.

$$P(\tau_0) = \frac{c^5(\tau_0)}{4\sqrt{V^2 - c^2(\tau_0)}} \left(\frac{\partial f(z_0, \tau_0)}{\partial z_0}\right)^2. \tag{6.15}$$

The geometrical divergence of the rays R for the uniformly and linearly moving source looks like

$$R(x,y) = \frac{c'(\tau_0)}{\sqrt{V^2 - c^2(\tau_0)}} \int_{\tau_0(x,y)}^{\tau(x,y)} c(t)\,dt + \sqrt{V^2 - c^2(\tau_0)} \tag{6.16}$$

Using (6.15), (6.16), it is possible to make out the evident mode of the first term of Airy wave at the motion of the point source of a mass in a non-stationary medium::

$$w = \frac{c^{5/2}(\tau_0)\alpha^2(x,y)f(z,\tau)}{2c^{3/2}(\tau)R^{1/2}(x,y)(V^2 - c^2(\tau_0))^{1/4}} \frac{\partial f(z_0,\tau_0)}{\partial z_0} \mathrm{Ai}'\!\left(\alpha(x,y)\frac{t - \tau(x,y)}{\varepsilon^{2/3}}\right) + O(\varepsilon^{4/3}).$$

For the numerical calculations they used the data on the variability of Brunt-Väisälä frequency in the real conditions of the Black sea. The Fig. 9 presents the profiles $N(z,t)$ with the time interval of 4 hours, for which the calculations were conducted. $N(z)$-distribution was used as the stationary and it is shown by the dashed line on the figure. The Fig. 10 and 11 present the results of the calculations of the first mode for the non-stationary (solid line) and the stationary (dashed line) cases at the following values of the parameters: $V = 1$ m/s, $\tau_0 = 0$, $\tau = 5000$ s, $z = -10$ м, $z_0 = -20$ м. The Fig. 10 presents results of calculations of the rays and wave fronts; the Fig. 11 presents the vertical velocity w in the fixed point of space T.

Thus, the solution of the problem on propagation of Airy and Fresnel internal waves in the stratified non-stationary medium and conducted on the basis of this problem solution numerical calculations demonstrate, that the variability of Brunt-Väisälä frequency in time can distinctly affect on the way of propagation of the internal gravity waves. Therefore at solution of the problem of propagation of the internal waves alongside with vertical stratification and dependence of $N^2$ on the horizontal variables it is necessary to consider also the variability of the stratified medium parameters in a time.

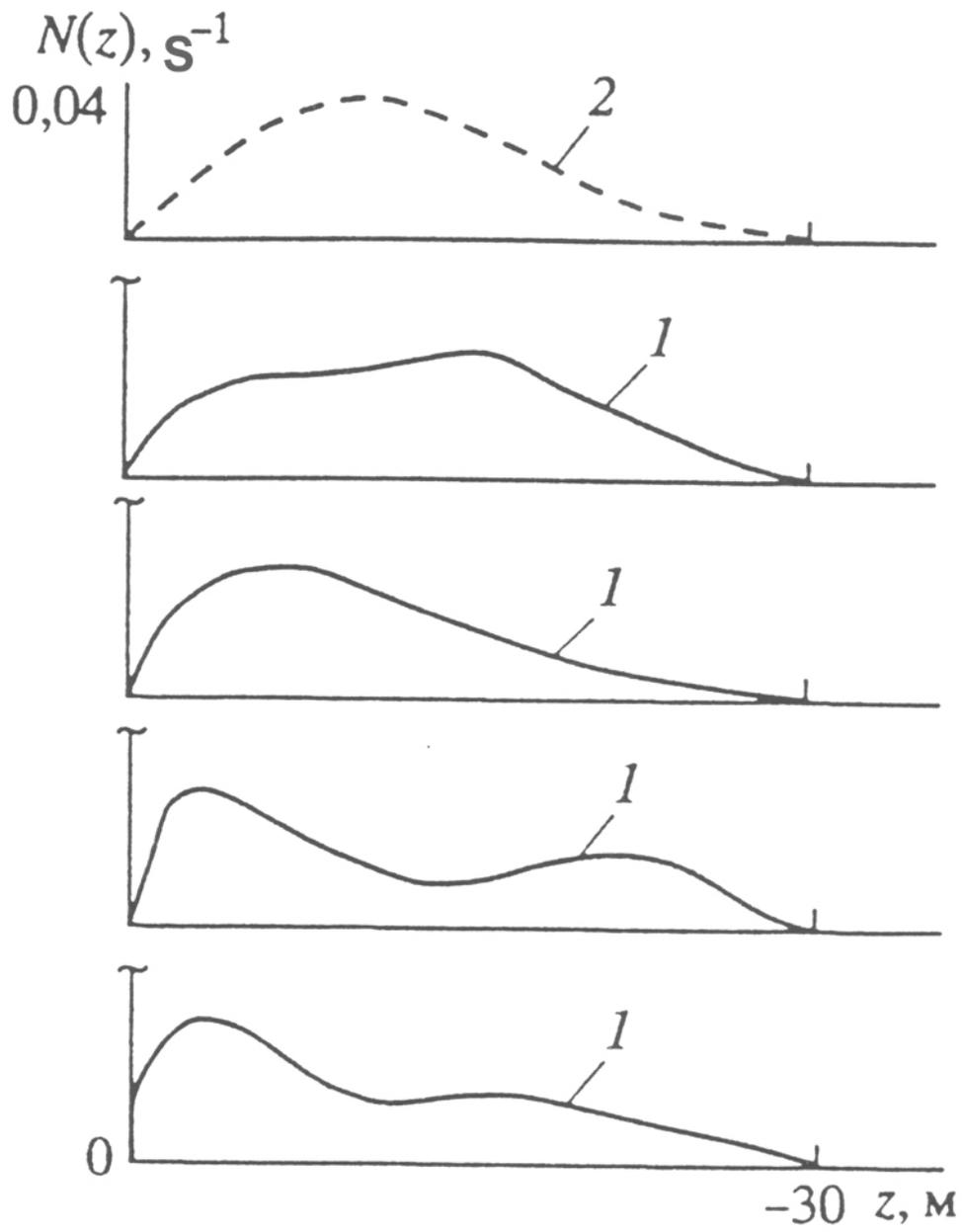

Fig.9

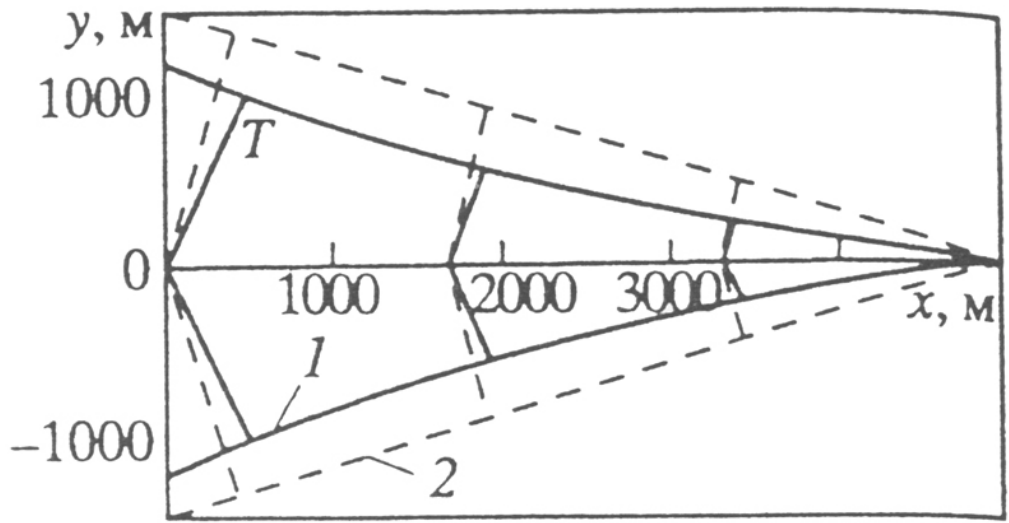

Fig.10

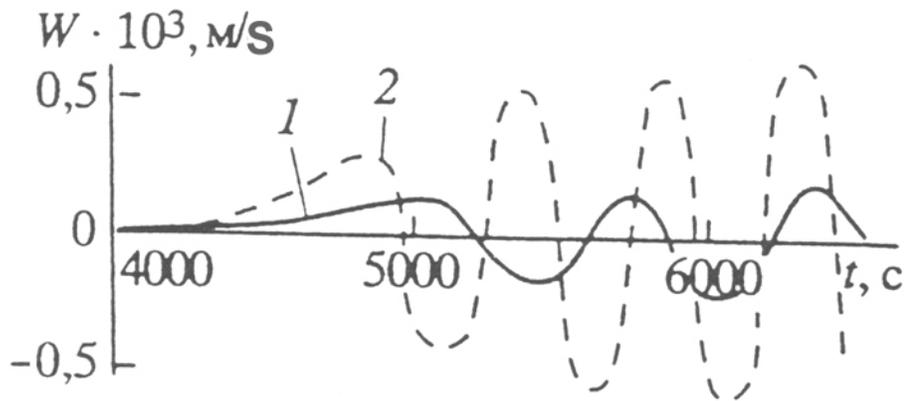

Fig.11